\documentclass[11pt]{article}

\usepackage{lineno}
\usepackage[hidelinks]{hyperref}
\usepackage{color}
\usepackage{amssymb,amsmath}
\usepackage{subfigure, float, graphicx}
\usepackage[square,super,sort&compress]{natbib}
\usepackage[margin=1in]{geometry}
\graphicspath{{Figs/}}
\usepackage[title,titletoc,toc]{appendix}

\modulolinenumbers[5]

\renewcommand{\vec}[1]{\ensuremath\boldsymbol{#1}}

\newcommand{\CA}{\ensuremath\textrm{Ca}}
\newcommand{\RE}{\ensuremath\textrm{Re}}
\newcommand{\RM}{\ensuremath\textrm{Rm}}
\newcommand{\MN}{\ensuremath\textrm{Mn}}
\begin{document}
	
\begin{center}
	\huge
	Vesicles in magnetic fields
	
	\bigskip

	\normalsize
	David Salac \\
	Department of Mechanical and Aerospace Engineering, \\ University at Buffalo SUNY, \\ 318 Jarvis Hall, Buffalo, NY 14260-4400, USA\\716-645-1460\\\href{mailto:davidsal@buffalo.edu}{davidsal@buffalo.edu}
\end{center}

\noindent\makebox[\linewidth]{\rule{\textwidth}{0.4pt}}
\bigskip

	Liposome vesicles tend to align with an applied magnetic field. This is due to the directional magnetic susceptibility difference
		of the lipids which form the membrane of these vesicles. In this work a model of liposome vesicles exposed to magnetic 
		field is presented. Starting from the base energy of a lipid membrane in a magnetic field, the force applied
		to the surrounding fluids is derived. This force is then used to investigate the dynamics of vesicle in the presence
		of magnetic fields.
\bigskip
\noindent\makebox[\linewidth]{\rule{\textwidth}{0.4pt}}

%%%%%%%%%%%%%%%%%%%%		End ARXIV     %%%%%%%%%%%%%%%%%%%%%%%%%%%%%%%%%%%%%%%%%%%%%%%%%%%%%%%%

\section{Introduction}

	Knowledge about the directed motion of biological and bio-compatible nano- and microstructures, particularly liposome vesicles, 
	is critically important for a number of biotechnologies. 
	For example,
	in directed drug delivery it is critical that drug carriers be directed towards locations where they are to release an encapsulated 
	drug.~\cite{Kagan2010} The directed motion of biological cells could be used to sort cells~\cite{Toner2005,Xia2006,Chen208} and to form 
	larger structures.~\cite{Solovev2010}
	Numerous techniques have been proposed as possible ways to control the motion of these soft-matter systems. Specially designed microfluidic 
	devices can use differences in size,~\cite{Tan2008} shape,~\cite{DuBose2014} and rigidity~\cite{Wang2013} to physically separate particles. 
	It is also possible to use light~\cite{MacDonald2003} and acoustic waves~\cite{Zhou2015} to induce particle motion.
	
	Of particular interest is the use of externally controllable fields to direct the motion of soft particles. Electric fields 
	have been demonstrated to induce large deformation in liposome vesicles both experimentally~\cite{Dimova2009,dimova2007} and 
	theoretically.~\cite{Kolahdouz2015a,salipante2014,schwalbe2011,mcconnell2013}
	Electric fields can be used to sort cells~\cite{Barrett2005,Cummings2003,Vahey2008} and to induce the formation of pores in 
	vesicle membranes.~\cite{Riske2005}
	
\begin{figure}
		\centering
		\includegraphics[width=6cm]{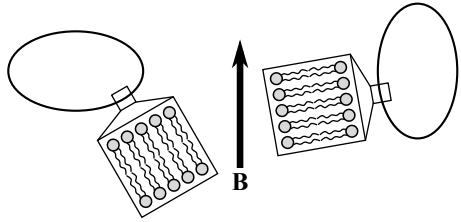}
		\caption{To minimize the total energy lipids will orient perpendicularly to an applied magnetic field. This results in a
		force which aligns the vesicle to the applied field.}		
		\label{fig:lipidAlignment}
	\end{figure}

	Magnetic fields also offer the opportunity to direct the motion of biologically compatible soft-matter.
	Experiments have demonstrated that liposome vesicles tend to align with an externally applied magnetic field.~\cite{BOROSKE1978}
	The phospholipids which compose the vesicle membrane are known to be diamagnetically anisotropic and tend to align 
	perpendicularly to an external field.~\cite{Rikken2014} Due to the nature of the liposome membrane this results in a rotational/alignment
	force which aligns and stretches a vesicle parallel to the applied field, see Fig. \ref{fig:lipidAlignment}.
	Due to the unique nature of vesicles this stretching is balanced by an increase in surface tension and bending energy.
	Using this fact Helfrich developed a model for the deformation of a vesicle when exposed to a magnetic field to 
	determine the flexibility of the vesicle membrane.~\cite{Helfrich1973,Helfrich1973a}
	
	Unlike electric field effects, the influence of magnetic fields on liposome vesicles has received much less attention. In addition to the 
	work of Helfrich mentioned above, it has been experimentally shown that liposomes made from dipalmitoylphosphatidyl choline (DPPC) 
	have temperature dependent deformation and permeability when exposed to magnetic fields.~\cite{Tenforde1988} In the same work a simple model
	was developed to explore the observed behavior. The fusion of liposome vesicles for a range of applied magnetic fields has also been
	experimentally demonstrated~\cite{Ozeki2000} while others have verified the alignment
	of vesicles to the external magnetic field.~\cite{Qiu1993,Kiselev2008}
	Theoretical and computational investigation of magnetohydrodynamics of liposome vesicles are much less common. Helfrich investigated
	the birefringence of vesicles in magnetic fields~\cite{Helfrich1973} and very few investigations into the biomechanics of vesicles in magnetic field
	have been performed.~\cite{Ye2015}
	
	To the author's knowledge this is the first effort to model the general dynamics of vesicles when exposed to externally driven magnetic fields.
	In the remainder of this work the governing equations and numerical methods used to model this system will be presented. Sample results of a 
	vesicle in the presence of magnetic fields will also be shown.
	
	%Look at [Ye and Curcur] Vesicle biomechanics in a time-varying magnetic. Eq. (2) indicates that the induced electric field will be zero for a constant electric field.

\section{Governing Equations}

\begin{figure}
		\centering
		\includegraphics[width=6cm]{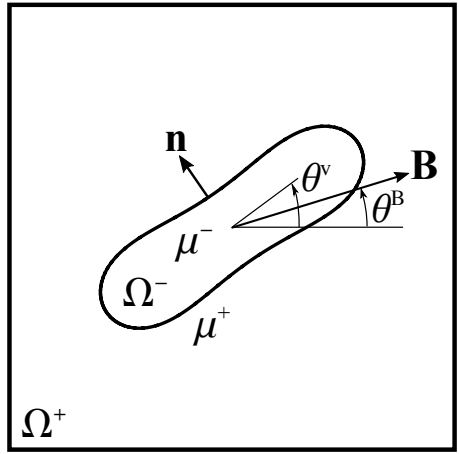}
		\caption{Schematic of a vesicle exposed to an externally driven magnetic field. The magnetic field $\vec{B}$ is at an angle of $\theta^B$ from the $x$-axis
		while the inclination angle of the vesicle is given by $\theta^v$. The inner, $\mu^-$, and outer, $\mu^+$, fluid viscosity may differ. The outward
		unit normal vector to the interface is given by $\vec{n}$.
		}
		\label{fig:schematic}
	\end{figure}

	Consider a vesicle suspended in a fluid and exposed to an externally driven magnetic field, Fig. \ref{fig:schematic}. 
	The density, electrical and magnetic properties between the inner and outer fluid are matched while the viscosity may vary. The magnetic
	field in static in time but could be spatially varying. The size of the vesicle 
	is on the order of 10 $\mu$m while the thickness of the membrane is approximately 5 nm, which allows for the modeling of the membrane as a thin
	interface separating two fluids.~\cite{Seifert1997} The vesicle membrane is impermeable to fluid molecules and 
	the number of lipids in a vesicle membrane does not change over time
	while the surface density of lipids at room temperature is constant.~\cite{Seifert1997} These conditions result in an inextensible membrane with
	constant enclosed volume and global surface area, in addition to local surface area incompressibility.
	
	For any multiphase fluid system the time-scale associated with charges migrating towards the interface is given by the charge relaxation time, 
	$t_c=\epsilon/\sigma$, where $\epsilon$ and $\sigma$ are the the fluid permittivity and conductivity, respectively.~\cite{Saville1997,Melcher1969} 
	Typical values for these in experimental vesicles investigations
	are $\epsilon\approx 10^{-9}$ F/m and $s\approx 10^{-3}$ S/m.~\cite{Riske2006,salipante2014,needham1989electro} This results in a charge relaxation time 
	of $t_c\approx 10^{-6}$ s, which is much faster than observed dynamics of vesicles when exposed to magnetic fields.~\cite{BOROSKE1978}
	It is thus valid to assume that there are no free charges in the bulk fluids and thus the leaky-dielectric model can be assumed.~\cite{Saville1997} 
	This lack of free charges in the bulk fluids has implications when considering the forces on the fluid. If there are free-charges in the fluid then 
	the Lorentz force will drag the fluid into motion, which is common for magnetohydrodynamics using conducting fluids.~\cite{Guo2013,Ki2010,Tagawa2006}
	In the absence of free charges the Lorentz force can be ignored in the bulk fluid and therefore fluid will be driven into motion only
	by conditions at the membrane. 
	
	In general the applied magnetic field and induced electric field are coupled through Maxwell's equations. Under the assumption
	that the magnetic field is static in time the only possible coupling between the electric field and the magnetic field is through the 
	electric current density, $\vec{j}=\sigma\left(-\nabla\Phi+\vec{u}\times\vec{B}\right)$, where $\Phi$ is the 
	induced electric potential, $\vec{u}$ is the fluid velocity and $\vec{B}$ is the applied magnetic field. As the induced electrical 
	current must divergence free, $\nabla\cdot\vec{j}=0$, 
	any induced electric potential will obey $\nabla\cdot\left(\sigma\nabla\Phi\right)=\nabla\cdot\left(\sigma\vec{u}\times\vec{B}\right)$.
	Experimental investigations of vesicles in a 1.5 T magnetic field demonstrate that responses take on the order of 10 s.~\cite{BOROSKE1978}
	Assuming that the distance traveled during this time is 20 $\mu$m this results in a velocity of $2\times10^{-6}$ m/s. In the
	absence of an external electric field this results in an induced electric current density of $3\times10^{-9}$ A/m$^2$.
	At such small induced current densities the induced electric field will be much smaller than those needed to induce vesicle 
	deformation.~\cite{Dimova2009,Kolahdouz2015a,salipante2014,vlahovska2009electrohydrodynamic} 
	From this analysis the induced electric field and it's contribution to the dynamics of the vesicle will be ignored.

\subsection{Forces Exerted by the Membrane}
	\label{sec:membraneForces}
	
	The motion of the fluid will be driven by the conditions and forces at the vesicle membrane. Let the lipid membrane be given by $\Gamma$.
	The total energy of the membrane is composed of four components:
	\begin{align}
		E[\Gamma] &= E_k[\Gamma] + E_\gamma[\Gamma] + E_{m,bulk}[\Gamma] + E_{m,rot}[\Gamma], \label{eq:membraneEnergy} \\
		E_k[\Gamma] &= \oint_\Gamma\left(\dfrac{1}{2}k_c H^2+k_g K\right)\;dA, \label{eq:bendingEnergy}\\
		E_\gamma[\Gamma] &= \oint_\Gamma \gamma\;dA, \label{eq:tensionEnergy}\\
		E_{m,bulk}[\Gamma] &= -\dfrac{\chi_\perp d}{2\mu_m}\oint_\Gamma B^2\;dA, \label{eq:magneticBulkEnergy} \\
		E_{m,rot}[\Gamma] &= -\dfrac{\Delta\chi d}{2\mu_m}\oint_\Gamma\left(\vec{n}\cdot\vec{B}\right)^2\;dA. \label{eq:magneticRotEnergy}		
	\end{align}
	The first integral, $E_k[\Gamma]$, provides the bending energy associated with the current membrane configuration where 
	$\kappa_{c}$ is the bending rigidity, $\kappa_{g}$ is the Gaussian bending rigidity, $K$ is the Gaussian curvature,
	and $H$ is the total curvature, which equals the sum of the principle curvatures. In this work the Gaussian curvature energy
	is ignored as the integral of the Gaussian curvature around any closed surface is a constant.~\cite{Carmo1976}
	The second integral, $E_\gamma[\Gamma]$, is the energy associated with a non-uniform 
	tension, $\gamma$. These two energies are based on the Helfrich model in the absence of spontaneous curvature\cite{Helfrich1973a} and have been used extensively to
	model liposome vesicles.~\cite{biben2003,salipante2014,vlahovska2007,Sohn2010} 
	
	The final two integrals, $E_{m,bulk}[\Gamma]$ and $E_{m,rot}[\Gamma]$, provide the energy of a lipid membrane in a magnetic field $\vec{B}$ 
	when the outward unit normal to the interface is given by $\vec{n}$ and $\vec{B}\cdot\vec{B}=B^2$.~\cite{Rikken2014,SCHOLZ1984} 
	The first energy, $E_{m,bulk}$, provides the total (bulk) energy of a membrane with a magnetic susceptibility perpendicular to a 
	lipid axis given by $\chi_\perp$, a membrane thickness of $d$, and where $\mu_m$ is the magnetic permeability of the membrane.
	As lipid molecules are diamagnetic materials, $\chi_\perp<0$.
	The second, $E_{m,rot}$, is the magnetic alignment energy, with $\Delta \chi=\chi_\parallel-\chi_\perp$ being 
	the difference between the magnetic susceptibilities in the parallel and 
	perpendicular direction for lipid molecules. 
	For lipid molecules, the 
	perpendicular magnetic susceptibility is larger (less negative) than the parallel one and thus $\Delta\chi<0$, 
	although it is possible to change this by adding biphenyl moieties to 
	the phospholipids.~\cite{Tan2002} This magnetic susceptibility difference drives the lipid molecules to become perpendicular to an applied magnetic field,
	which in turn causes the lipid vesicle itself to align with the field.~\cite{BOROSKE1978}
	
	In order to minimize the energy, the membrane will exert a force on the surrounding fluid. These forces are calculated by taking the variation of the appropriate membrane
	energy with respect to a change of membrane location.
	For a vesicle and neglecting spontaneous curvature the ultimate forms of the bending and tension forces are found to be\cite{seifert1999fluid}
	\begin{align}
		\vec{\tau}_{k} &= -\kappa_c\left(\frac{1}{2}H^3-2HK + \nabla^2_{s}H\right)\vec{n},\\
		\vec{\tau}_{\gamma} &=\gamma H\vec{n}-\nabla_s\gamma.
	\end{align}
	The surface gradient, $\nabla_s$, and surface Laplacian, $\nabla_s^2$, are defined using the projection operator 
	$\vec{P}=\vec{I}-\vec{n}\otimes\vec{n}$. More precisely, the surface gradient of a scalar field
	is given by $\nabla_s f=\vec{P}\nabla f$ while the surface Laplacian is $\nabla_s^2 f=\nabla_s\cdot\nabla_s f$.

	The magnetic force has not been presented in the literature and is derived here. In particular, the method outlined by 
	Napoli and Vergori is used to determine the variation of the energy.~\cite{Napoli2010} Let a generic energy functional be given by 
	$\oint_\Gamma w\;dA$, where $w$ is an arbitrary energy functional density per unit area which only a function of 
	the unit normal $\vec{n}$ and no other surface quantities. The first variation of the energy
	with respect to a change of the interface is then given by $\nabla_s\cdot\left(w\vec{P}-\vec{n}\otimes\left(\vec{P}\vec{w}_n\right) \right)$,
	where $\vec{w}_n=\partial w/\partial\vec{n}$ is the derivative of the energy density with respect to the unit normal. 
	
	Assume that a single lipid species is present. Therefore the material properties $\chi_\perp$, $\Delta \chi$, $d$, and $\mu_m$ are all
	constants. First consider the contribution of the bulk energy of the lipid membrane in a magnetic field, $\oint_\Gamma B^2 dA$.
	In this case $w=B^2$ and thus $\vec{w}_n=0$. Using the results shown in Appendix \ref{App:Geometric}, this results in
	\begin{equation}
		\nabla_s\cdot\left(B^2 \vec{P}\right) = \nabla_s B^2 - B^2 H \vec{n}.
	\end{equation}
	
	Next consider the rotational energy contribution, $\oint_\Gamma\left(\vec{n}\cdot\vec{B}\right)^2 dA$. From $w=\left(\vec{n}\cdot\vec{B}\right)^2=B_n^2$
	the quantity $\vec{w}_n=2\left(\vec{n}\cdot\vec{B}\right)\vec{B}=2 B_n\vec{B}$ is obtained. Thus,
	\begin{equation}
		\nabla_s\cdot\left(B_n^2\vec{P}-\vec{n}\otimes\left(2B_n\vec{P}\vec{B}\right)\right) 			
			= \nabla_s\cdot\left(B_n^2\vec{P}\right)-2\nabla_s\cdot\left(\vec{n}\otimes\left(B_n\vec{P}\vec{B}\right)\right)
			\label{eq:magForceDerivation0}
	\end{equation}	
	where $B_n=\vec{n}\cdot\vec{B}$ is the portion of the magnetic field in the normal direction. 
	Each of these components are considered in turn. The first component results in
	\begin{equation}
		\nabla_s\cdot\left(B_n^2\vec{P}\right) = \nabla_sB_n^2 - B_n^2 H \vec{n} = 2B_n\nabla_s B_n-B_n^2 H\vec{n}.
		\label{eq:magForceDerivation1}
	\end{equation}
	The second component is 
	\begin{align}
		\nabla_s\cdot\left(\vec{n}\otimes\left(B_n\vec{P}\vec{B}\right)\right)
			&=B_n \left(\nabla_s\vec{n}\right)\vec{P}\vec{B}+\vec{n}\nabla_s\cdot\left(B_n\vec{P}\vec{B}\right) \nonumber \\
			&=B_n \left(\nabla_s\vec{n}\right)\vec{B} + \vec{n}\left(\vec{B}\cdot\nabla_s B_n-B_n^2 H+B_n\nabla_s\cdot\vec{B}\right) \nonumber \\
			&=B_n \vec{L}\vec{B} + \vec{n}\left(\vec{B}\cdot\nabla_s B_n-B_n^2 H+B_n\nabla_s\cdot\vec{B}\right),
			\label{eq:magForceDerivation2}
	\end{align}
	where the surface gradient of the unit normal, $\nabla_s\vec{n}=\vec{L}$, is called the curvature tensor, or shape operator, 
	of the interface. It is a symmetric and real matrix which characterizes the curvature of the surface.~\cite{Fried2008, Napoli2010, Napoli2012}
	One eigenvalue of $\vec{L}$ is zero and has a corresponding eigenvector in the direction of the unit normal, $\vec{n}$.
	As $\vec{L}$ is symmetric and real, it can be decomposed as $\vec{L}=\kappa_t \vec{t}\otimes\vec{t}+\kappa_b \vec{b}\otimes\vec{b}$,
	where $\kappa_t$ and $\kappa_b$ are the remaining eigenvalues of $\vec{L}$ with corresponding eigenvectors $\vec{t}$ and $\vec{b}$, respectively. 
	In this case the eigenvalues are principle curvatures of the interface while the eigenvectors are the principle tangent directions.
	Thus, the first part of Eq. \eqref{eq:magForceDerivation2} can be written as
	\begin{align}
		B_n \vec{L}\vec{B} 
			= B_n\left(\kappa_t \vec{t}\otimes\vec{t}+\kappa_b \vec{b}\otimes\vec{b}\right)\vec{B}
			= B_n\left(\kappa_t B_t \vec{t} + \kappa_b B_b \vec{b}\right),
			\label{eq:magForceDerivation3}
	\end{align}
	where $B_t=\vec{t}\cdot\vec{B}$ and $B_b=\vec{b}\cdot\vec{B}$ are the components of the magnetic field in the principle directions.
	
	Combining the results of Eqs. \eqref{eq:magForceDerivation0}-\eqref{eq:magForceDerivation3} and simplifying results in
	\begin{align}
		&\nabla_s\cdot\left(B_n^2\vec{P}-\vec{n}\otimes\left(2B_n\vec{P}\vec{B}\right)\right) = \nonumber \\
			&2B_n \nabla_s B_n + B_n^2 H \vec{n} - 2 B_n\left(\kappa_t B_t \vec{t} + \kappa_b B_b \vec{b}\right) - 2\vec{n}\left(\vec{B}\cdot\nabla_s B_n\right)-2B_n \vec{n}\nabla_s\cdot\vec{B}.
	\end{align}
	
	Using these results the force due to the magnetic field is then
	\begin{align}		
		\vec{\tau}_{m,bulk} = &-\dfrac{\delta E_{m,bulk}}{\delta \Gamma} = \dfrac{\chi_\perp d}{2\mu_m}\left(\nabla_s B^2-B^2 H\vec{n}\right), \label{eq:magBulkForce}  \\				
		\vec{\tau}_{m,rot} = &-\dfrac{\delta E_{m,rot}}{\delta \Gamma} = \dfrac{\Delta\chi d}{2\mu_m}\left( 2B_n\nabla_s B_n + B_n^2 H \vec{n} \right.\nonumber \\
							& \left.- 2 B_n\left(\kappa_t B_t \vec{t} + \kappa_b B_b \vec{b}\right) - 2\vec{n}\left(\vec{B}\cdot\nabla_s B_n\right)-2B_n \vec{n}\nabla_s\cdot\vec{B}\right), \label{eq:magRotForce} 
	\end{align}
	recalling that $B^2=\vec{B}\cdot\vec{B}$, $B_n=\vec{n}\cdot\vec{B}$, $B_t=\vec{t}\cdot\vec{B}$, and $B_b=\vec{b}\cdot\vec{B}$.
	
	For general situations, the above expressions works well. When the magnetic field is spatially constant simplifications can be made by
	expanding the $\nabla_s B_n=\nabla_s\left(\vec{n}\cdot\vec{B}\right)$ terms:
	\begin{align}
		\nabla_s B_n=\nabla_s\left(\vec{n}\cdot\vec{B}\right)=\vec{B}\cdot\nabla_s\vec{n}+\vec{n}\cdot\nabla_s\vec{B}=\vec{B}\cdot\vec{L}+\vec{n}\cdot\nabla_s\vec{B}=\vec{B}\cdot\vec{L}=\vec{L}\vec{B},
	\end{align}
	as $\nabla_s\vec{B}=\left(\nabla\vec{B}\right)\vec{P}=\vec{0}$ and $\vec{L}=\vec{L}^T$.
	Beginning with Eq. \eqref{eq:magRotForce} and using the $\vec{L}\vec{B}$ form this results in
	\begin{align}
		\vec{\tau}_{m,rot} 
			=& \dfrac{\Delta\chi d}{2\mu_m}\left( 2B_n\nabla_s B_n + B_n^2 H \vec{n} - 2 B_n \vec{L}\vec{B} - 2\vec{n}\left(\vec{B}\cdot\nabla_s B_n\right)-2B_n \vec{n}\nabla_s\cdot\vec{B}\right) \nonumber \\
			=& \dfrac{\Delta\chi d}{2\mu_m}\left( 2B_n \vec{L}\vec{B} + B_n^2 H \vec{n} - 2 B_n \vec{L}\vec{B} - 2\vec{n}\left(\vec{B}\cdot\vec{L}\vec{B}\right)\right) \nonumber \\			
			=& \dfrac{\Delta\chi d}{2\mu_m}\left( B_n^2 H - 2\vec{B}\cdot\vec{L}\vec{B}\right) \vec{n} \nonumber \\
			=& \dfrac{\Delta\chi d}{2\mu_m}\left( B_n^2 H - 2\vec{B}\cdot\left(\kappa_t B_t\vec{t}+\kappa_b B_b\vec{b}\right)\right) \vec{n} \nonumber \\
			=& \dfrac{\Delta\chi d}{2\mu_m}\left( B_n^2 H - 2 \kappa_t B_t^2-2\kappa_b B_b^2\right) \vec{n},
	\end{align}
	due to the fact that $\nabla_s\cdot\vec{B}=\vec{P}:\nabla\vec{B}=0$ when $\vec{B}$ is spatially constant.

\subsection{Interface Description}
	
	In this work a level-set formulation is used to describe the location of the vesicle membrane. Let the evolving interface be given as the set of points where 
	a level-set function $\phi(\vec{x},t)$ is zero: $\Gamma(t)=\left\{\vec{x}:\phi(\vec{x},t)=0\right\}$, where $\vec{x}$ is a position in space and $t$ is time.
	Instead of explicitly tracking the location of the interface $\Gamma$ through time the position is implicitly tracked by evolving $\phi$.	
	Following convention the inner fluid, $\Omega^-$, is given by the region $\phi<0$ while the outer fluid, $\Omega^+$, is given by $\phi>0$. 
	The entire domain is denoted as $\Omega=\Omega^-\cup\Omega^+$. Using the 
	level-set geometric quantities are easily obtained. For example, the outward facing unit normal vector and the total curvature (sum of principle 
	curvatures) is given by
	\begin{align}
		\vec{n}=&\dfrac{\nabla\phi}{\|\nabla\phi\|}, \\
		H =& \nabla\cdot\vec{n}.
	\end{align}
	It is also possible to use the level set function to define varying material parameters using a single relation.
	Consider the determination of the viscosity at any location in the domain. Letting $\mu^-$ be the inner viscosity and 
	$\mu^+$ be the outer viscosity the viscosity at a point $\vec{x}$ is given by $\mu(\vec{x})=\mu^{-}+(\mu^{+}-\mu^{-})\mathcal{H}(\phi(\vec{x}))$,
	where $\mathcal{H}$ is the Heaviside function. Similar expressions hold for other material quantities.
	In practice a smoothed version of the Heaviside function is used to ensure numerical stability.~\cite{Towers2009}
	Finally, motion of the interface is obtained by advecting the level set function,
	\begin{equation}
		\dfrac{\partial \phi}{\partial t}+\vec{u}\cdot\nabla\phi=0.
		\label{eq:levelsetadvection}
	\end{equation}
	Details of the numerical implementation will be presented later.

\subsection{Fluid Flow Equations}
		
	Define the bulk fluid hydrodynamic stress tensor in each fluid as
	\begin{equation}
		 \vec{T}_{hd}^{\pm}=-p^{\pm}\vec{I}+\mu^{\pm}(\nabla\vec{u}^{\pm}+\nabla^T\vec{u}^{\pm}) \textnormal{  in  } \Omega^{\pm}.
		\label{eq:hydrodynamic_stress}
	\end{equation}	
	The forces derived in Section \ref{sec:membraneForces} are balanced by a jump in the fluid stress tensor,
	\begin{equation}
		\vec{n}\cdot(\vec{T}_{hd}^+-\vec{T}_{hd}^-) = \vec{\tau}_{k} + \vec{\tau}_{\gamma} + \vec{\tau}_{m,bulk} + \vec{\tau}_{m,rot}.
		\label{eq:interface_force}
	\end{equation}	
	Note that in general there is a contribution from a jump in the Maxwell stress tensor acting on the interface. In the absence of electric fields and with
	matched magnetic fluid properties this contribution is zero and thus is not included.
		
	Using the level set formulation it is possible to write the fluid momentum equations and the interface force balance as a single equation valid over the
	entire domain:\cite{Chang1996,Kolahdouz2015b}
	\begin{align}		
		\rho(\phi)\frac{D \vec{u}}{Dt}=&-\nabla p + \nabla\cdot\left(\mu(\phi)\left(\nabla\vec{u}+\nabla^{T}\vec{u}\right)\right) \nonumber \\
			&+\delta(\phi)\kappa_{c}\left(\frac{H^3}{2} -2KH+\nabla^2_s H \right)\nabla\phi \nonumber \\
			&+\delta(\phi)\|\nabla\phi\|\left(\nabla_s\gamma-\gamma H\vec{n}\right) \nonumber \\
			&-\delta(\phi)\|\nabla\phi\|\dfrac{\chi_\perp d}{2\mu_m}\left(\nabla_s B^2-B^2 H\vec{n}\right) \nonumber \\
			&-\delta(\phi)\|\nabla\phi\|\dfrac{\Delta\chi d}{2\mu_m}\left( 2B_n\nabla_s B_n + B_n^2 H \vec{n} \right.\nonumber \\
				&\left.- 2 B_n\left(\kappa_t B_t \vec{t} + \kappa_b B_b \vec{b}\right) 
				 - 2\vec{n}\left(\vec{B}\cdot\nabla_s B_n\right)-2B_n \vec{n}\nabla_s\cdot\vec{B}\right),			
		\label{eq:CSF_model}
	\end{align}
	where the full form of the force, Eqs. \eqref{eq:magBulkForce} and \eqref{eq:magRotForce}, have been used.
	The use of the Dirac function $\delta(\phi)$ localizes the contributions from the membrane forces near the $\phi=0$ contour, which is the location of the interface.\cite{Towers2008}
	Volume and surface area conservation are provided by ensuring that 
	\begin{align}	
		\nabla\cdot\vec{u}&=0 \textrm{  in  } \Omega,\\
		\nabla_s\cdot\vec{u}&=0 \textrm{  on  } \Gamma.
	\end{align}
	Note that in the fluid formulation, Eq. (\ref{eq:CSF_model}), the tension is used to enforce surface area constraint $\nabla_s\cdot\vec{u}=0$ and is computed
	as part of the problem alongside the pressure.
		
\subsection{Nondimensional Parameters and Equations}
	
	Assume that the density is matched between the inner and outer fluids while the viscosity has a ratio of $\eta=\mu^-/\mu^+$. Each of the forces 
	acting on the vesicle membrane have an associated time-scale which depends on the material properties. The time-scale associated
	with the bending of the membrane is\cite{schwalbe2011} 
	\begin{equation}
		t_k=\dfrac{\mu^+(1+\eta) l_0^3}{\kappa_c},
	\end{equation}
	where $l_0$ is the characteristic length scale.
	
	The magnetic field introduces two times scales, one associated with each component of the magnetic 
		field energy, Eqs. \eqref{eq:magneticBulkEnergy} and \eqref{eq:magneticRotEnergy}.
	In both case the magnetic forces are compared to the viscous forces.
	The first is the time scale of the bulk energy,
	\begin{equation}
		t_{m,bulk}=\dfrac{\mu^+(1+\eta)l_0 \mu_m}{|\chi_\perp| d B_0^2},
	\end{equation}
	while the magnetic rotation time scale is
	\begin{equation}
		t_{m,rot}=\dfrac{\mu^+(1+\eta)l_0 \mu_m}{|\Delta\chi| d B_0^2},
	\end{equation}
	where $B_0$ is the characteristic magnetic field strength and recalling
	that $\mu_m$ is the magnetic permeability of the membrane. 
	
	Let the characteristic time be given by $t_0$, which allows for the definition of the characteristic velocity: $u_0=l_0/t_0$. 
	Define a capillary-like number providing the relative strength of the bending, $\CA=t_k/t_0$, 
	a magnetic Mason number indicating the strength of the bulk motion, $\MN=\textrm{sgn}(\chi_\perp)t_{m,bulk}/t_0$,
	and a magnetic field induced rotational force number, $\RM=\textrm{sgn}(\Delta \chi)t_{m,rot}/t_0$,
	while the Reynolds number is given by $\RE=\rho u_0 l_0/\mu^+=\rho\;l_0^2/(\mu^+ t_0)$.
	The use of $\textrm{sgn}(\chi_\perp)$ and $\textrm{sgn}(\Delta \chi)$ 
	takes into account the fact that the values of $\chi_\perp$ and $\Delta \chi$ can be either positive or negative. Therefore,
	the dimensionless parameters $\MN$ and $\RM$ can either be positive or negative, depending on if the membrane is a paramagnetic
	or diamagnetic material.	
	It is then possible to write
	the dimensionless fluid equations as
	\begin{align}		
		\frac{D \vec{u}}{Dt}=&-\nabla p + \dfrac{1}{Re}\nabla\cdot\left(\mu(\phi)\left(\nabla\vec{u}+\nabla^{T}\vec{u}\right)\right) \nonumber \\
			&+\dfrac{1}{\CA\;\RE}\delta(\phi)\left(\frac{H^3}{2} -2KH+\nabla^2_s H \right)\nabla\phi \nonumber \\
			&+\delta(\phi)\|\nabla\phi\|\left(\nabla_s\gamma-\gamma H\vec{n}\right)\nonumber \\
			&-\dfrac{1}{2\;\RE\;\MN}  \delta(\phi)\|\nabla\phi\|\left(\nabla_s B^2-B^2 H\vec{n}\right) \nonumber \\
			&-\dfrac{1}{2\;\RE\;\RM} \delta(\phi)\|\nabla\phi\|\left(2 B_n \nabla_s B_n + B_n^2 H \vec{n} - 2 B_n\left(\kappa_t B_t \vec{t} + \kappa_b B_b \vec{b}\right) \right.\nonumber \\
				& \hspace{1cm} \left.- 2\vec{n}\left(\vec{B}\cdot\nabla_s B_n\right)-2B_n \vec{n}\nabla_s\cdot\vec{B}\right),
		\label{eq:CSF_nondim_model}
	\end{align}	
	where all quantities are now dimensionless and the viscosity at a point can be calculated using $\mu(\phi)=\eta+(1-\eta)\mathcal{H}(\phi)$.
	
	The energy is normalized by using the bending rigidity, $\kappa_{c}$, as the characteristic energy scale. When writing the normalized total energy,
	the contribution from the tension, Eq. \eqref{eq:tensionEnergy}, is not included as it is a co-dimension one parameter used to enforce surface incompressibility.
	Therefore, the normalized energy of the system is then
	\begin{equation}
		E[\Gamma] = \dfrac{1}{2} \oint_\Gamma H^2 dA - \dfrac{1}{2}\dfrac{\CA}{\MN}\oint_\Gamma B^2 dA - \dfrac{1}{2}\dfrac{\CA}{\RM}\oint_\Gamma\left(\vec{n}\cdot\vec{B}\right)^2 dA.
	\end{equation}
	
	It is useful to compare the time-scales associated with a typical experiment. Boroske and Helfrich reported the alignment of vesicles in 
	magnetic fields with a strength of 1.5 T and reported that the magnetic susceptibility difference to be $-3.52\times10^{-8}$ in SI units,
	assuming a membrane thickness of 6 nm.~\cite{BOROSKE1978}	
	Assuming the viscosity is matched and equal to water, $\mu^+=10^{-3}$ Pa s, the membrane magnetic permeability is that of free-space, $\mu_m=4\pi\times10^{-7}$ H/m,
	and a characteristic length of 10 $\mu$m (as estimated by the figures in Boroske and Helfrich) the membrane rotation time is $t_{m,rot}\approx 53$ s. It was
	reported by Boroske and Helfrich that rotation through an angle of $\pi/2$ took approximately 100 s, which matches well with the characteristic time calculated here.
	Using the same values and given that the bending rigidity for the system is approximately $10^{-19}$ J,\cite{Beblik1985} the time-scale associated with bending is 20 s, which 
	agrees with the experimental results as no deformation of the membrane was observed during rotation.
	
\section{Numerical Methods}
	\label{sec:numericalMethods}
	Two numerical methods must be discussed. The first is the advection of the level set field. In this work a new semi-implicit level set Jet scheme is used.~\cite{Velmurugan2016}
	In addition to the level set function the gradient of the level set are also tracked, which increases the accuracy of the method.~\cite{Seibold2012}
	The extension allows the original level set Jet scheme to be used for stiff advection problems. It is composed of three main steps. First, the level set field 
	is advanced using a second-order, semi-implicit, semi-Lagrangian update,
	\begin{equation}
		\dfrac{3\phi^{n+1}-2\phi_d^n+\phi_d^{n-1}}{2\Delta t}=\beta\nabla^2\phi^{n+1}-\beta\nabla^2\hat{\phi},
	\end{equation}
	where $\beta=0.5$ is a constant, $\phi_d^n$ is the departure value of the level set at time $t^n$, 
	$\phi_d^{n-1}$ is the departure value of the level set at time $t^{n-1}$, and $\hat{\phi}=2\phi^n-\phi^{n-1}$ is an approximation of the level set value at time
	$t^{n+1}$. Once the smooth level set field is obtained the 
	effect of smoothing is captured by defining a source term,
	\begin{equation}
		S_{\phi}=\beta\nabla^2\left(\phi^{n+1}-\hat{\phi}\right).
	\end{equation}
	This advection source term is used to update the level set values on a sub-grid which surrounds all grid points,
	\begin{equation}
		\dfrac{3\phi^{s,n+1}-2\phi_d^{s,n}+\phi_d^{s,n-1}}{2\Delta t}=S_{\phi}.
	\end{equation}
	Using these updated sub-grid level set values, $\phi^{s,n+1}$, finite difference approximations are used to calculate the updated gradient field.
	It was shown that this method results in an accurate
	and stable scheme for the modeling of moving interfaces under stiff advection fields.~\cite{Velmurugan2016}
	
	The fluid field is obtained using a projection-based method.~\cite{Kolahdouz2015b} 
	The first step is to calculate a tentative field using a semi-implicit, semi-Lagrangian
	method:
	\begin{align}
		\dfrac{3\vec{u}^{\ast}-2\vec{u}_d^n+\vec{u}_d^{n-1}}{2\Delta t}=&-\nabla \hat{p} +\dfrac{1}{\RE}\nabla\cdot\left(\mu\left(\nabla \vec{u}^{\ast} +\nabla^T\hat{\vec{u}}\right)\right)\nonumber \\
			&+\vec{f}_k^n+\vec{f}_\gamma^n+\vec{f}_{m,bulk}^n+\vec{f}_{m,rot}^n,
	\end{align}
	where $\vec{f}_k^n$, $\vec{f}_\gamma^n$ and $\vec{f}_{m,bulk}^n+\vec{f}_{m,rot}^n$ are the bending, tension, and magnetic forces 
	while $\vec{u}_d^n$ and $\vec{u}_d^{n-1}$ are the departure velocities and $\hat{p}=2p^n-p^{n-1}$ is an extrapolation of the pressure 
	to time $t^{n+1}$.
	The next step is to calculate the corrections to the pressure and tension to enforce volume and surface area conservation,
	\begin{equation}
		\dfrac{3}{2}\dfrac{\vec{u}^{n+1}-\vec{u}^{\ast}}{\Delta t}=-\nabla q +\delta(\phi)\|\nabla\phi\|\left(\nabla_s \xi-\xi H\nabla\phi \right),
		\label{eq:projection}
	\end{equation}
	where $q$ and $\xi$ are the corrections needed for the pressure and tension, respectively. The pressure and tension are computed simultaneously 
	to enforce both local and global conservation of the enclosed volume and surface area. Complete details of the method are provided in
	Kolahdouz and Salac.~\cite{Kolahdouz2015b}
		
\section{Results}

	The experimental results most applicable are those of Boroske and Helfrich.~\cite{BOROSKE1978} As such the results presented here 
	will be modeled after those experiments. The characteristic time is chosen to be $t_0=1$ s. Assuming a bending rigidity of 
	$\kappa_c\approx 25 k_B T \approx 10^{-19}$ J,\cite{Vitkova2013} a vesicle radius of $10\;\mu$m, fluid density of 1000 kg/m$^3$, and outer fluid viscosity of $10^{-3}$ Pa s
	with matched viscosity ($\eta=1$) the capillary-like number becomes $\CA=20$ while the Reynolds is $\RE=10^{-4}$.
	 
	Using a magnetic susceptibility difference on the order of $\Delta \chi=-3\times10^{-8}$ in SI units, with a membrane 
	magnetic permeability of $\mu_m=4\pi\times 10^{-7}$ H/m and membrane thickness of 6nm, the magnetic rotation constant
	scales as $\RM\sim -140/B_0^2$. Assuming a magnetic field strength between $B_0=1$ T and $B_0=10$ T, this results in a range of 
	$-150\lesssim \RM \lesssim -1$. 
	
	The magnetic susceptibility perpendicular to the lipid axis, $\chi_\perp$, is not 
	readily available in the literature, and thus it is not clear what the magnitude should be.	
	In spatially constant magnetic field the bulk magnetic energy contribution, Eq. \eqref{eq:magneticBulkEnergy}, is constant
	for incompressible membranes and thus the bulk magnetic force, Eq. \eqref{eq:magBulkForce}, does not need to be included.
	For spatially varying magnetic fields it is assumed that the magnetic susceptibility perpendicular is of the same order as the 
	magnetic susceptibility difference and therefore the magnetic Mason number will be taken to be $-100\lesssim \MN \lesssim -1$
	
	Due to the long simulation times, and to facilitate a larger number of trials,
	the results will be in the two-dimensional regime, resulting in zero Gaussian curvature: $K=0$.
	Unless otherwise stated, the computational domain is a square spanning $[-6.4,6.4]^2$ using a $257^2$ grid and 
	periodic boundary conditions while the time step is fixed at $\Delta t=0.1 h$, where $h=12.8/256=0.05$ is the grid spacing.
	The choice of this domain size and time step is justified in Sec. \ref{sec:domainParameters}.
	
	Vesicles are be characterized by several parameters. Specifically, the viscosity ratio $\eta=\mu^-/\mu^+$, the 
	inclination angle $\theta^v$, the deformation parameter $D$, and the reduced area $\nu$. 
	Inclination angles are determined by calculating the eigenvalues and eigenvectors of
	the vesicle's inertia tensor about its center of mass. The eigenvector corresponding to
	the larger of the two eigenvalues provides the direction of the long axis of a vesicle. The angle
	between the eigenvector associated with the long axis of the vesicle and the $x$-axis is denoted as the inclination angle.
	The deformation parameter is given by $D=(a-b)/(a+b)$, where $a$ and $b$ are the long and short axes of an ellipse with 
	the same inertia tensor as the vesicle.~\cite{Salac2012,RAMANUJAN1998}
	The vesicle reduced area indicates how deflated a vesicle is compared to a circle with the same interface length, and is given by $\nu=4 A\pi/L^2$, where
	$A$ and $L$ are the enclosed area and interface length, respectively. A value of $\nu=0.5$ indicates that the enclosed area is one-half of a circle with the same
	interface length while $\nu=1$ denotes a circle. 	
	
	All simulations begin with an ellipse having an interfacial length of $2\pi$ and a reduced area of $\nu=0.71$.
	This reduced area was estimated from Boroske and Helfrich, Fig. 1.~\cite{BOROSKE1978}
	The vesicle	is then allowed to evolve in the absence of a magnetic field to obtain a shape near the bending energy minimum.~\cite{Salac2011} This 
	shape is then used as the initial condition for the magnetically driven results.
	The initial orientation of all vesicles is vertical, which is denoted as having an inclination angle of $\theta^v=\pi/2$, see Fig. \ref{fig:schematic}.
	It is also assumed throughout that the viscosity is matched,  $\eta=1$.

\subsection{Direct comparison with Boroske and Helfrich}

\begin{figure}[]
		\centering
		\includegraphics[width=8cm]{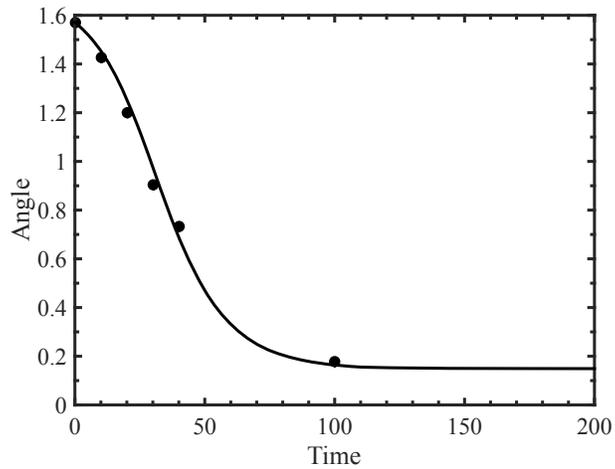}
		\caption{The inclination angle as a function of time for a vesicle with reduced area of $\nu=0.71$ in a magnetic field at an angle of $0.045\pi$ with
			a rotation constant of $\RM=-7.5$. The dots indicate the angles determined from Fig. 1 of Boroske and Helfrich,\cite{BOROSKE1978} after an appropriate rotation is done
			to take into account the different initial angles.}
		\label{fig:boroskeAngle}
	\end{figure}

\begin{figure}
		\centering
		\includegraphics[width=8cm]{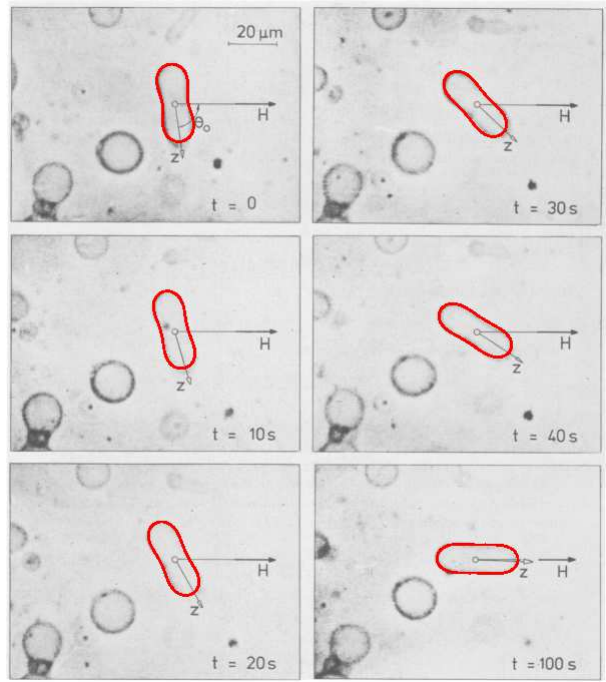}
		\caption{Comparison between experimental results of Boroske and Helfrich and the simulation. Due to the different initial orientation, the simulation
		results are first flipped about the horizontal axis and then rotated $0.045\pi$ counter-clockwise.
		Reprinted from Biophysical Journal, Vol 24 (3), Boroske and Helfrich, ``Magnetic anisotropy of egg lecithin membranes", Pages 863-868., December 1978, with permission from Elsevier.}
		\label{fig:boroske}
	\end{figure}

	To provide the reader a better understanding of the vesicle dynamics, a direct comparison with the results of Boroske and Helfrich is performed.~\cite{BOROSKE1978}
	Using Fig. 1 from that manuscript, it was estimated that the angle between the vesicle and the applied magnetic
	field is $0.455\pi$. In the simulation, 
	the vesicle is initially aligned with the vertical axis and the magnetic field has an angle of $\theta^B=0.045\pi$, which matches 
	the conditions of the experiment. As the magnetic field is spatially constant the bulk magnetic energy is ignored while the dimensionless magnetic
	rotation constant is set to $\RM=-7.5$. The inclination angle up to a time of 200 is shown in Fig. \ref{fig:boroskeAngle} while the computationally
	derived vesicle shapes are compared to the experimental result in Fig. \ref{fig:boroske}.

	The computational results match very well with the experimental results. Assuming a membrane thickness of $d=6$ nm, the properties of water, and an
	applied magnetic field strength of $B_0=1.5$ T, and using the value of $\RM=-7.5$, the magnetic susceptibility difference is calculated
	to be $\Delta \chi=-2.48\times 10^{-7}$. While this value is larger than that estimated by Boroske and Helfrich, it is within other experimentally
	determined values.~\cite{BRAGANZA1984}
	
\subsection{Verification of domain parameters}
\label{sec:domainParameters}

	To verify the choice of domain size, grid size, and time step a systematic investigation is performed by varying each simulation parameter
	individually. 
	The magnetic field is spatially constant and fixed at an angle of $\theta^B=0.045\pi$. Three magnetic rotation constants used are $\RM=-1$,
	$\RM=-10$, and $\RM=-100$. As the magnetic field is spatially constant, the bulk magnetic field contribution is neglected.
	
\begin{figure}
		\centering
		\includegraphics[width=8cm]{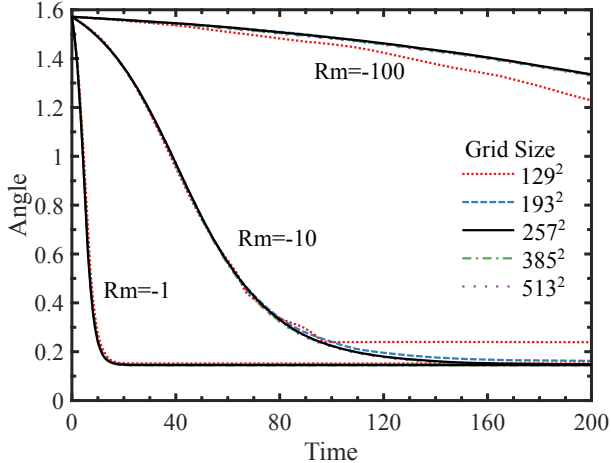}
		\caption{The inclination angle versus time for magnetic rotation strengths of $\RM=-1$, $\RM=-10$, and $\RM=-100$ and grid sizes ranging from $129^2$ to $513^3$. 
			All results use a domain size of $[-6.4,6.4]^2$ with a time step of $\Delta t=0.1h$, where $h$ is the grid spacing. 
			No change in the results are seen past a grid size of $257^2$.}		
		\label{fig:domainN}
	\end{figure}

	First consider the influence of the grid size on the results. Using a $[-6.4,6.4]^2$ domain, grid sizes ranging from $129^2$ to $513^2$ 
	are used. In all cases the time step is set to $\Delta t = 0.1h$, where $h$ is the grid spacing. The results shown in Fig. \ref{fig:domainN}
	indicate that a grid size of $129^2$ is not sufficient. This shouldn't be surprising as with this grid spacing only approximately 8 grid 
	points are used to describe the vesicle at it's narrowest point. The difference in the results using more than a grid size of $257^2$ are not noticeable 
	for any of the three magnetic rotation constants, which justifies that particular choice.
	
\begin{figure}
		\centering
		\includegraphics[width=8cm]{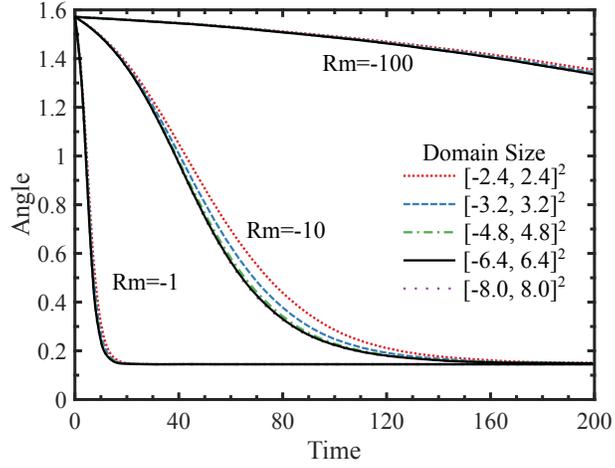}
		\caption{The inclination angle versus time for magnetic rotation strengths of $\RM=-1$, $\RM=-10$, and $\RM=-100$ and domain sizes ranging from 
			$[-2.4,2.4]^2$ to $[-8.0,8.0]^2$. The number of grid points is adjusted so that a constant grid spacing of $h=0.05$ and 
			constant time step $\Delta =0.1h$ is used for each simulation.
			No change in the results are seen past a domain size of $[-6.4,6.4]^2$.}
		\label{fig:domainL}
	\end{figure}

	Next consider the influence of domain size on the rotation dynamics. Using a constant grid spacing of $h=0.05$ and time step of $\Delta t=0.1h$,
	various domain sizes from $[-2.4,2.4]^2$ to $[-8.0,8.0]^2$ are considered, see Fig. \ref{fig:domainL}. Clearly, boundary effects are 
	present in the smallest domains, particularly when $\RM=-10$. Once the domain size reaches $[-6.4, 6.4]^2$, only small differences are observed.	
	
	Finally consider the influence of the time step on the rotation dynamics. Using a $[-6.4,6.4]^2$ domain with $257^2$ grid points, various time
	steps from $\Delta t=0.02h$ to $\Delta t=0.5h$ are considered. Note that using time steps of $\Delta t=h$ proved unstable.
	There are almost no differences using time steps smaller than $\Delta t=0.1h$, and thus that is the time step chosen for further results.
	
	\begin{figure}
		\centering
		\includegraphics[width=8cm]{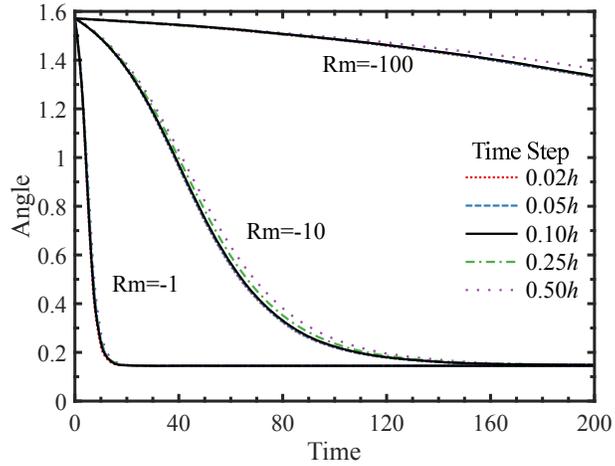}
		\caption{
			The inclination angle versus time for magnetic rotation strengths of $\RM=-1$, $\RM=-10$, and $\RM=-100$ and time steps ranging from 
			$\Delta t=0.02 h$ to $\Delta t=0.5h$. The domain is fixed at $[-6.4,6.4]^2$ while the size of the domain is $257^2$. 					
			No change in the results are seen past a time step of $\Delta t=0.1h$.
		}
		\label{fig:domainDT}
	\end{figure}

\subsection{Influence of $\RM$}
	\label{sec:RM}

\begin{figure}
		\centering
		\subfigure[Inclination angle over time for various values of $\RM$, indicated by the numbers.]{
			\includegraphics[width=8cm]{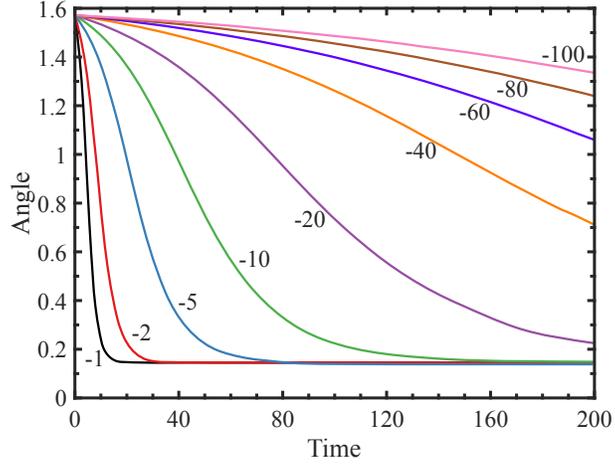}
			\label{fig:RmResults_Angle}
		}		
		\subfigure[Time it takes to rotate through an angle of $0.05\pi$, $0.25\pi$, and $0.4\pi$.]{
			\includegraphics[width=8cm]{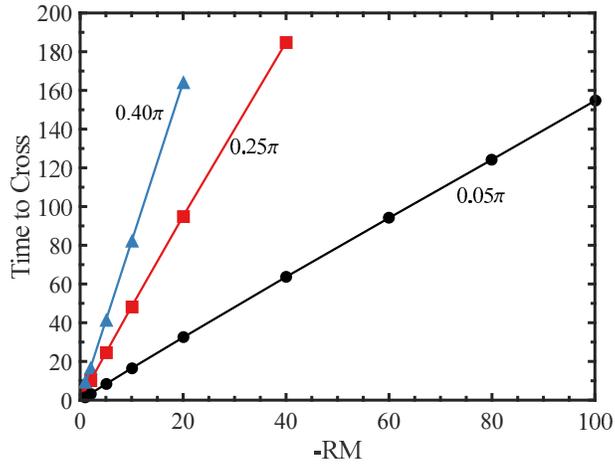}
			\label{fig:RmResults_TOC}
		}		
		\caption{The influence of the magnetic field-induced rotational force, $\RM$, on the inclination angle.
			As $\RM$ increases it takes additional time to align with the magnetic field.}
		\label{fig:RmResults}
	\end{figure}

	The influence of the magnetic rotation force is explored by varying $\RM$ within the range from 1 to 100
	up to a time of $t=200$. The resulting inclination angle over time is shown in Fig. \ref{fig:RmResults_Angle},
	while the amount of time needed to rotate through an angle of $0.05\pi$, $0.25\pi$, and $0.4\pi$
	is shown in Fig. \ref{fig:RmResults_TOC}. There are several points to be made. First,
	the equilibrium angle of the vesicle, given enough time, will match that of the applied magnetic field. 
	Second, the amount of time that is required to rotate through a particular angle is linearly dependent on the 
	$\RM$ value. It should be noted that due to the definition of $\RM$, this is related to the quadratic of the magnetic 
	field strength, i.e. a 2-fold increase in the magnetic field results in a 4-fold decrease of the $\RM$ parameter.
	Therefore, increasing the magnetic field strength by a factor of two reduces the amount of time needed to rotate by a factor
	of four.
	
\begin{figure}
		\centering
		\subfigure[$\RM=-1$]{\label{fig:RM_energy_A} \includegraphics[width=8cm]{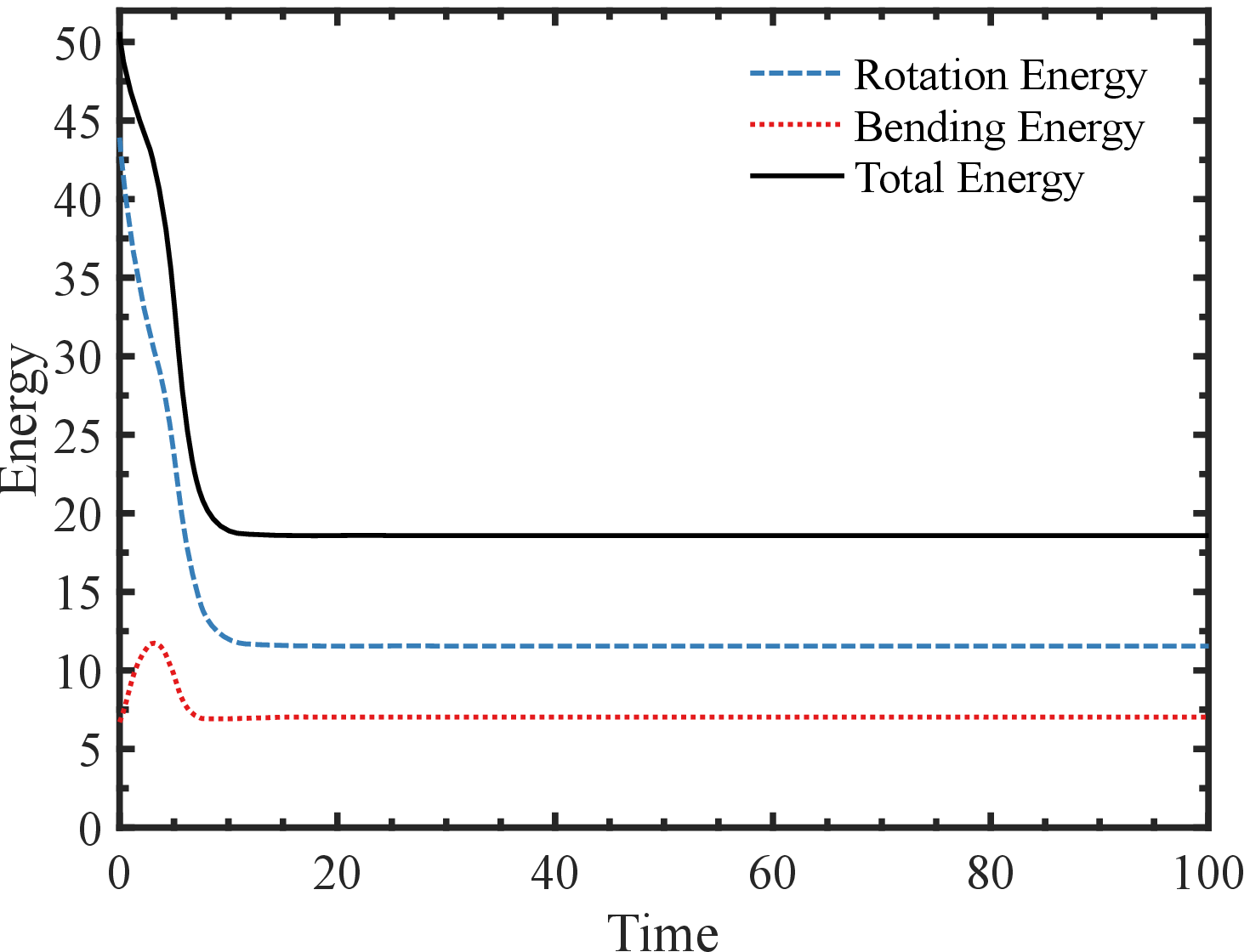}}
		\subfigure[$\RM=-10$]{\label{fig:RM_energy_B} \includegraphics[width=8cm]{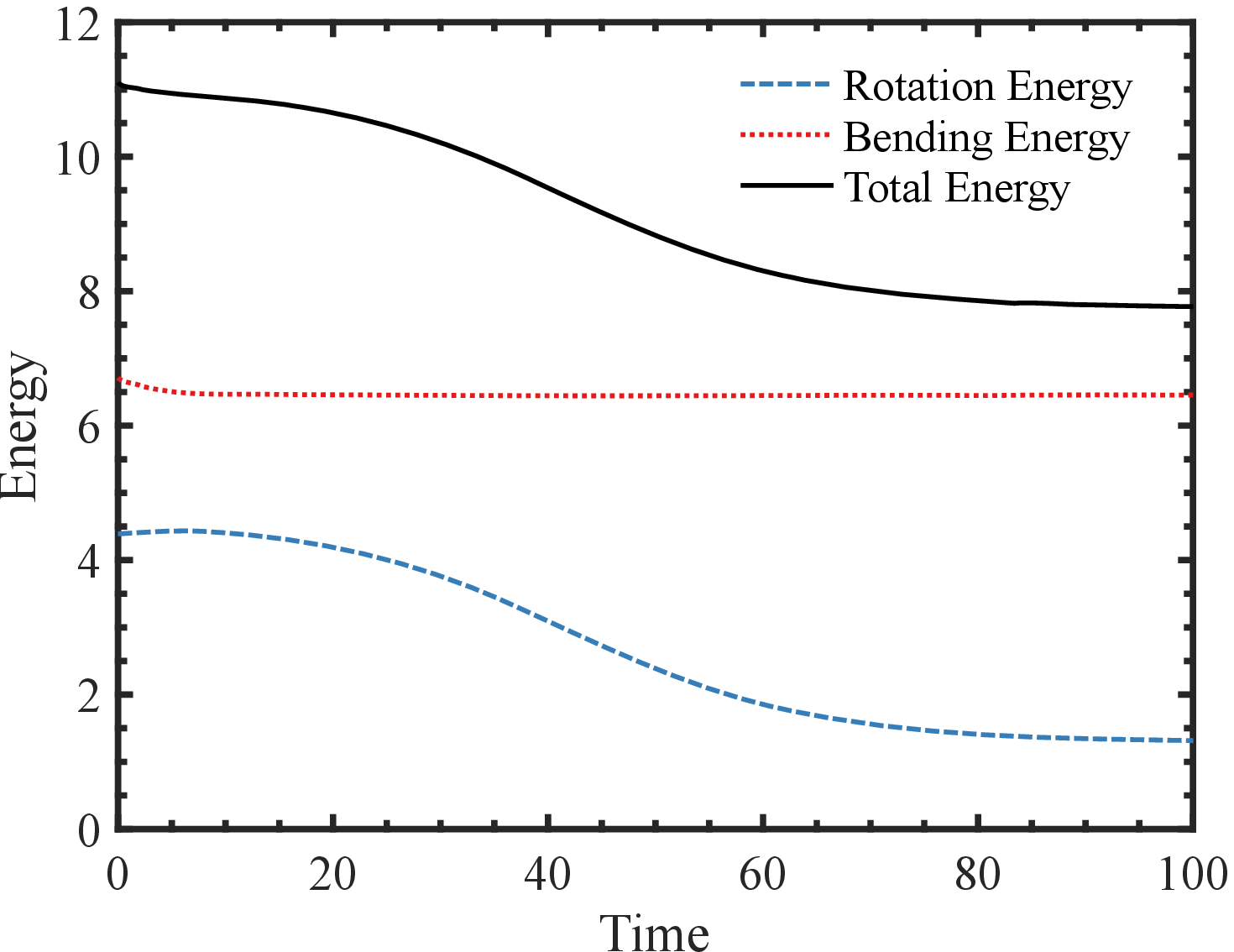}}
		\subfigure[$\RM=-100$]{\label{fig:RM_energy_C} \includegraphics[width=8cm]{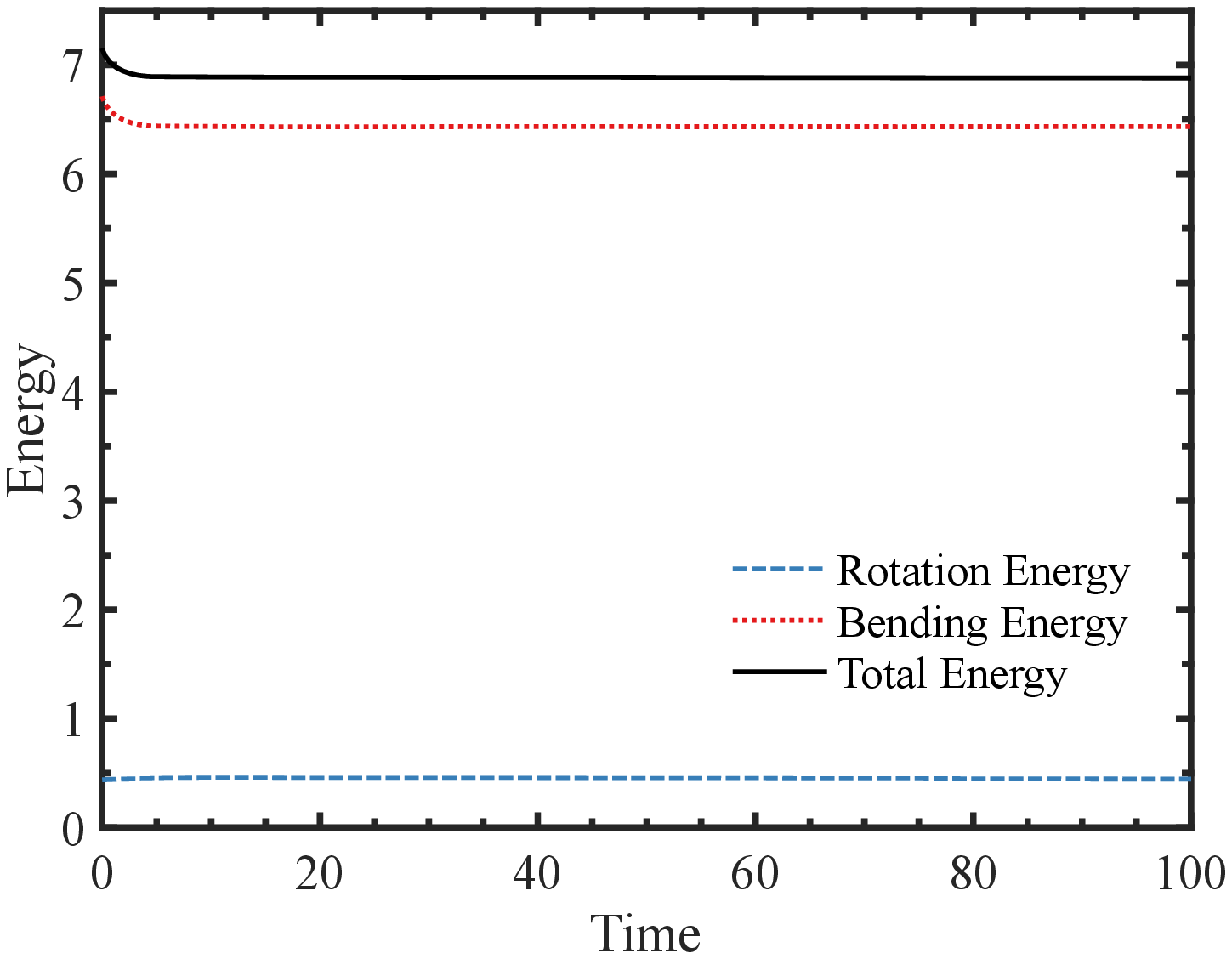}}
		\caption{The bending, rotation, and total energy for rotation strengths of $\RM=-1$, $\RM=-10$, and $\RM=-100$.}
		\label{fig:RM_energy}
	\end{figure}

	An investigation of the energy for three characteristic rotation strengths, $\RM=-1$, $\RM=-10$, and $\RM=-100$, is shown in Fig. \ref{fig:RM_energy}.
	As expected, the total energy decreases as the vesicle becomes aligned with the magnetic field. The overall rotation rate is directly correlated
	to the initial rotation energy, as a higher initial magnetic rotation energy correlates to faster rotation time. It is interesting to note that when
	the magnetic rotation strength is strong, $\RM=-1$, the vesicle membrane can not respond quickly to changes in bending energy and thus the 
	bending energy contribution increases, Fig. \ref{fig:RM_energy_A}. This is in contrast to the weaker rotation forces shown in 
	Figs. \ref{fig:RM_energy_B} and \ref{fig:RM_energy_C}, where both the rotation and bending energy are strictly decreasing. 

\begin{figure}
		\centering
		\includegraphics[width=6cm]{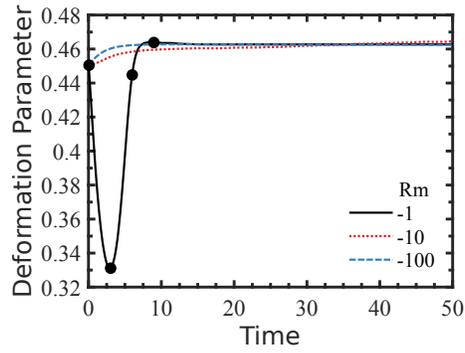}
		\caption{
			The deformation parameter for $\RM=-1$, $\RM=-10$, and $\RM=-100$. Strong magnetic field effects induce larger shape deformations. 
				The circles on the $\RM=-1$ correspond to the interfaces shown in Fig. \ref{fig:RM001_Shapes}. The shapes for $\RM=-10$ and $\RM=-100$ 
				do not look qualitatively different from that shown in Fig. \ref{fig:boroske}.
		}
		\label{fig:RM_DeformationParameter}
	\end{figure}

\begin{figure}
		\centering
		\subfigure[$t=0$]{\includegraphics[width=4cm]{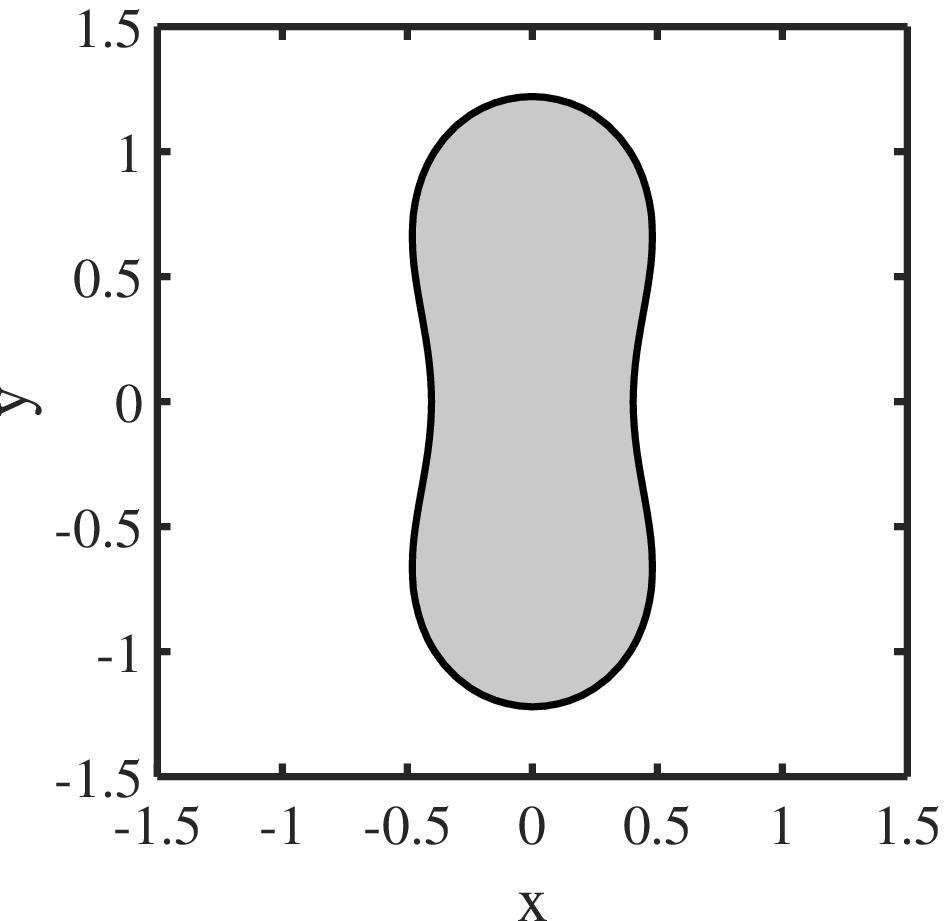}}
		\vspace{0.5cm}
		\subfigure[$t=3$]{\includegraphics[width=4cm]{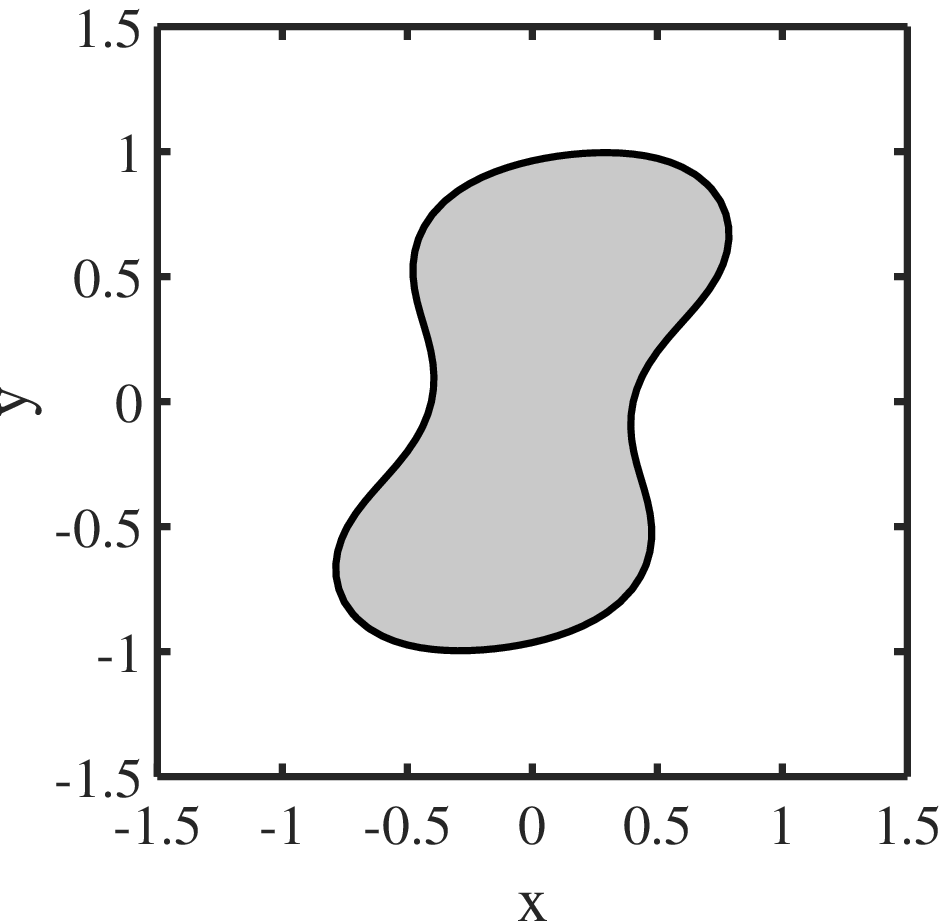}} \\
		\subfigure[$t=6$]{\includegraphics[width=4cm]{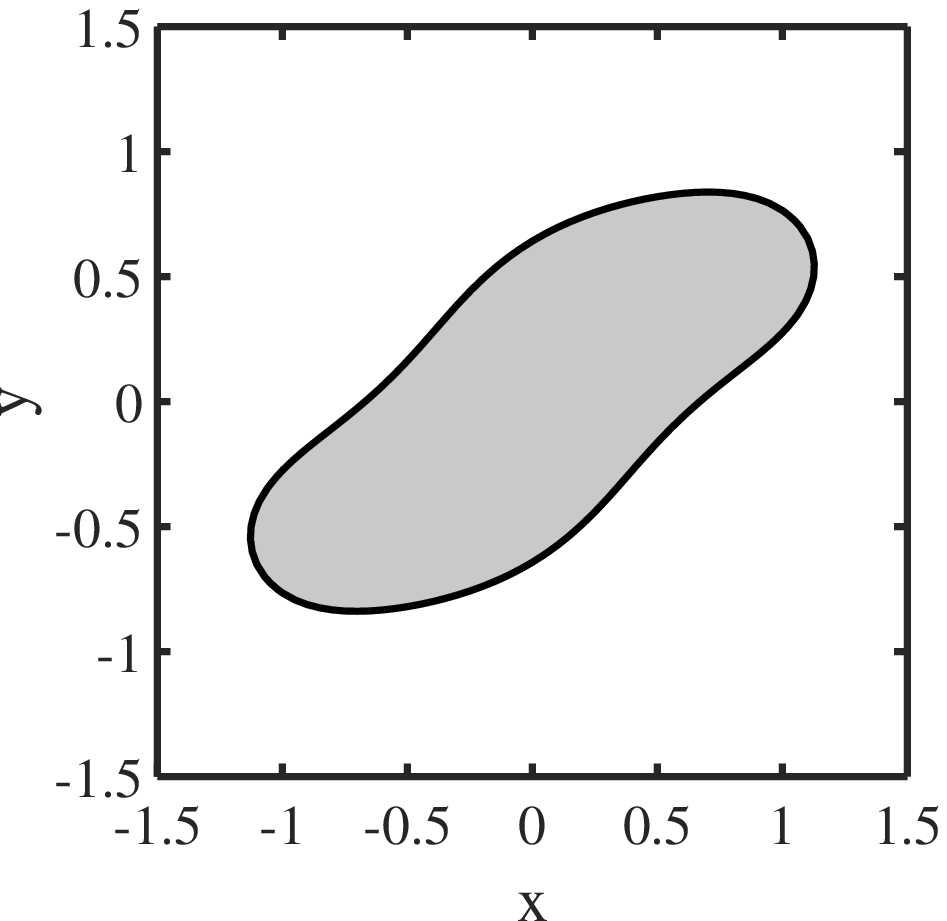}}
		\vspace{0.5cm}
		\subfigure[$t=9$]{\includegraphics[width=4cm]{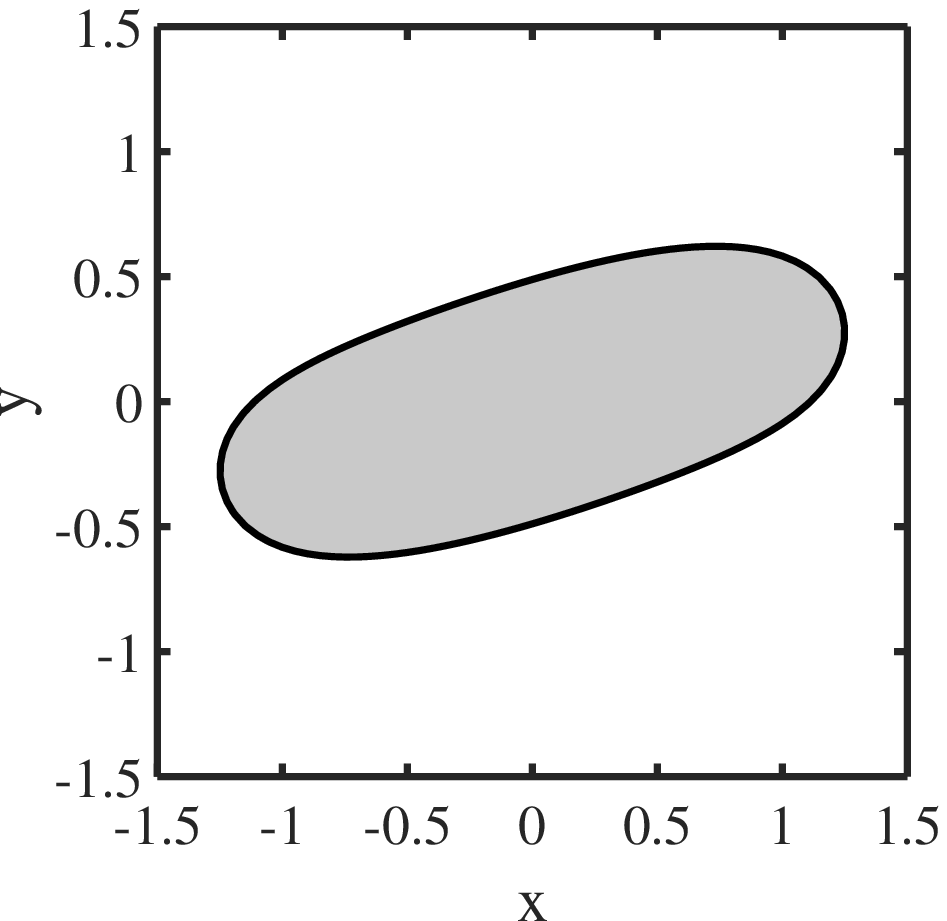}}
		\caption{The shape of the vesicle at various times using $\RM=-1$.}
		\label{fig:RM001_Shapes}
	\end{figure}

	To further explore the influence of $\RM$ on the vesicle shape, the deformation parameter for the three characteristic 
	$\RM$ values is shown in Fig. \ref{fig:RM_DeformationParameter}. It is clearly observed that the strong rotation force
	given by $\RM=-1$ causes larger deformations than the $\RM=-10$ and $\RM=-100$ cases. The shape of the vesicle using $\RM=-1$, as shown in
	Fig. \ref{fig:RM001_Shapes}, can be compared to that shown in Fig. \ref{fig:boroske}, and it is clear that larger deformation are observed
	before the vesicle flattens out.

\begin{figure}
		\begin{center}
			\subfigure[Interface Location at the times indicated]{\label{fig:Horizontal_Interface} \includegraphics[width=6cm]{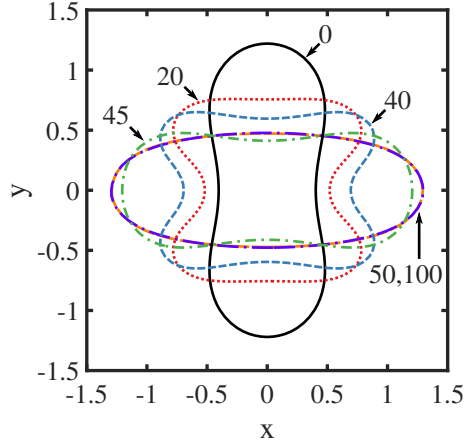}}
			
			\subfigure[Angle and Deformation Parameter]{\label{fig:Horizontal_AD} \includegraphics[width=8cm]{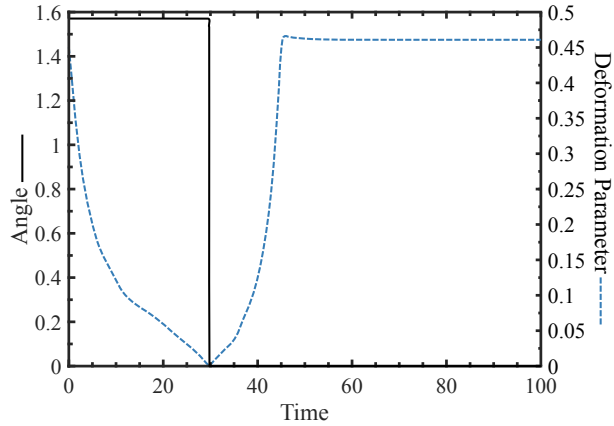}}
		\end{center}
		\caption{Sample interface locations, the inclination angle, and deformation parameter for a vesicle in a magnetic field aligned 
				along the $x$-direction. In this case $\RM=-1$.}
		\label{fig:Horizontal}
	\end{figure}

	It should be noted that the initial angle between the long-axis of the vesicle and the applied magnetic field is less than $\pi/4$. If the 
	angle is equal to $\pi/4$, then the mechanism of alignment is no longer rotation, but large-scale deformation of the interface.
	This can be seen in Fig. \ref{fig:Horizontal}, where the vesicle is placed in a magnetic field aligned with the $x$-axis and 
	the rotation/alignment strength is $\RM=-1$. Up until approximately $t=30$, the major axis of the vesicle is aligned with the $y$-axis.
	The vesicle undergoes large deformations, as demonstrated by the decrease in the deformation parameter. After $t=30$, the major axis is 
	aligned with the magnetic field along the $x$-axis and the vesicle begins to elongate to reduce both the bending and magnetic energies.
	It should be noted that the final deformation parameter for this case, $D\approx 0.46$, is similar to that shown in Fig. \ref{fig:RM_DeformationParameter},
	despite the difference in the magnetic angle.
	
\subsection{Spatially varying magnetic field}
	\label{sec:VaryingMagneticField}

\begin{figure}
		\centering
		\includegraphics[width=6cm]{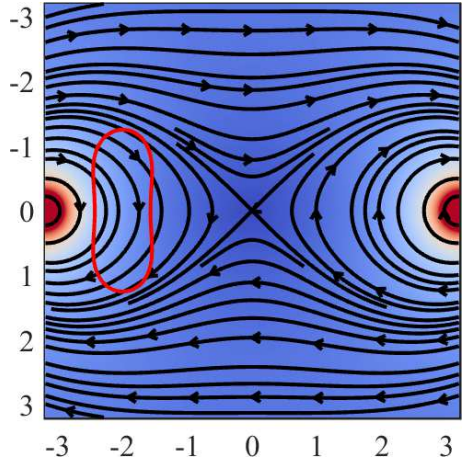}
		\caption{The magnetic field lines, initial location of the vesicle, and representation of the magnetic field strength (color online). The magnetic field is strongest 
			at the center of the two current carrying wires located at $(-3.2,0)$ and $(3.2,0)$, indicated by the color red, and quickly decays towards the center of the domain, indicated
			by blue.}
		\label{fig:magneticField}
	\end{figure}

	Next consider the influence of a spatially varying magnetic field. In this case, the full magnetic energy contribution must be considered, and thus both 
	$\RM$ and $\MN$ will be varied. To construct the variable magnetic field, the vesicle is placed inside a domain spanning $[-3.2,3.2]^2$ using a 
	grid size of $129^2$ so that $h=0.05$. In this case wall boundary conditions are assumed. To induce the magnetic field, two infinitely long 
	wires are placed at the locations $(-3.2,0)$ and $(3.2,0)$. Each of these wires has a current of magnitude $I_0$ in the vertical $z-$direction
	and produces a magnetic field given by 
	\begin{equation}
		\vec{B} = \dfrac{B_0}{2\pi}\left(\dfrac{x-a}{((x-a)^2+(y-b)^2},-\dfrac{y-b}{((x-a)^2+(y-b)^2} \right)
	\end{equation}
	where the $B_0=\mu_0 I_0$ is the strength of the induced magnetic field surrounding a wire at $(a,b)$.~\cite{Tzirtzilakis2005} Due to the linearity of the magnetic field,
	the total magnetic field is simply the summation of that induced by both wires. 
	A vesicle with a reduced area of $\nu=0.71$ and matched viscosity, $\eta=1$, is then centered at $(-2,0)$. It is expected that the vesicle will 
	migrate towards the center of the domain, which is the location of lowest magnetic field strength. 
	An example of the magnetic field and initial vesicle location is given in Fig. \ref{fig:magneticField}, which shows both the magnetic field lines and the intensity of the magnetic field.
	
\begin{figure*}
		\centering
		\includegraphics[width=12cm]{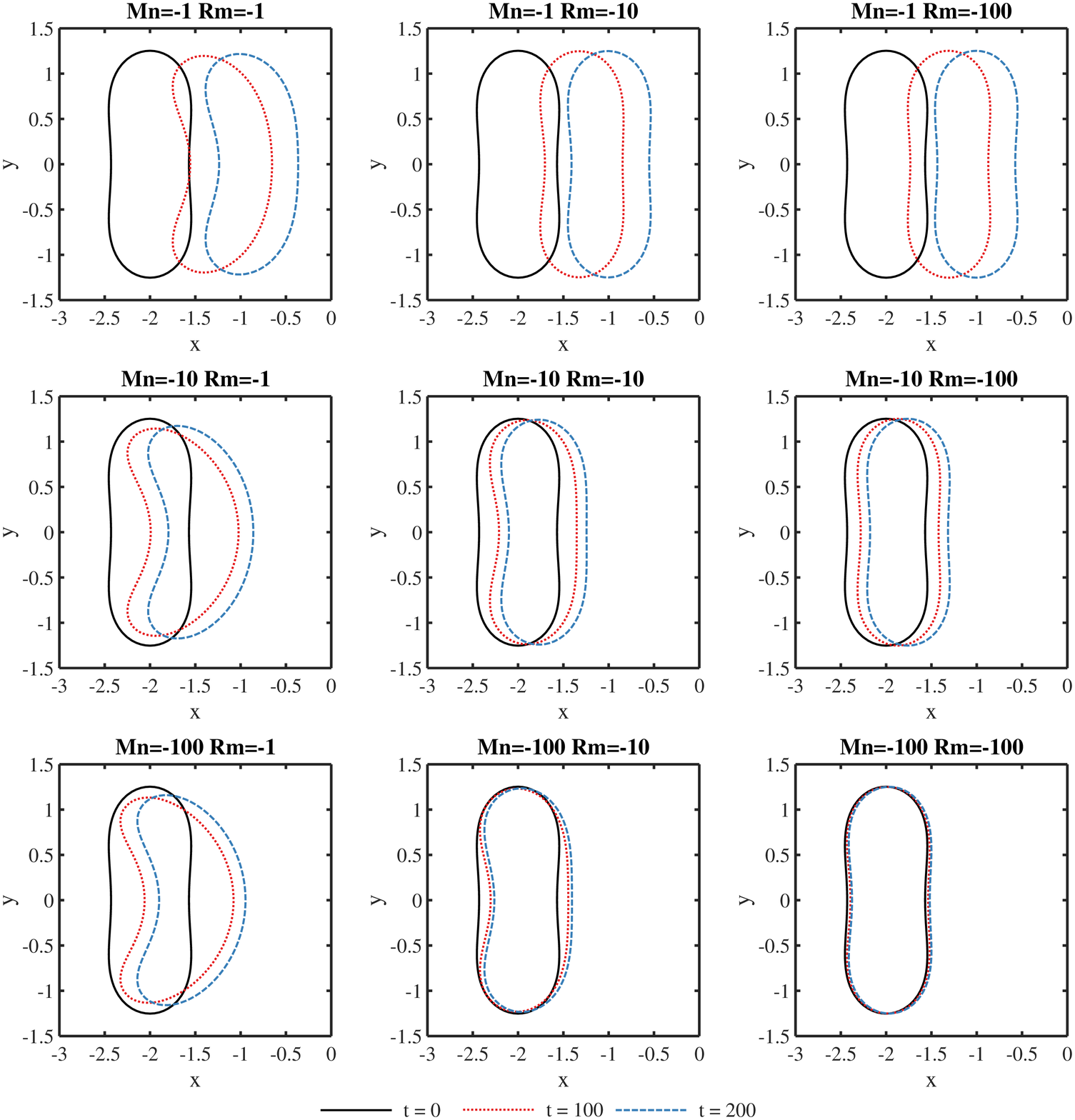}
		\caption{The location of the interface at times $t=0$, $t=100$, and $t=200$ for a spatially variable magnetic field using various values of $\MN$ and $\RM$.
			The results show a portion of the entire domain, which spans $[-3.2,3.2]^2$. The magnetic field arises due to current carrying wires embedded at locations
			$(-3.2,0)$ and $(3.2,0)$.}
		\label{fig:variableInterfaces}
	\end{figure*}

	The location of the interface at times of $t=0$, $t=100$, and $t=200$ for $\MN$ and $\RM$ values between 1 and 100 is shown in Fig. \ref{fig:variableInterfaces}. 
	In all cases the interface migrates towards the center of the domain. The rate of this migration and the overall deformation of the interface strongly depends 
	on both the $\MN$ and $\RM$ parameters. In general, as the strength of the alignment and bulk magnetic effects increases, the rate of of migration also 
	increases. It should also be observed that for stronger rotational strengths, denoted by lower $\RM$ values, the vesicle tends to align with the local magnetic 
	field. As the underlying local magnetic field is close to circular, the interface adopts this configuration.
	
\begin{figure}
		\centering
		\subfigure[$x$-centroid]{
			\includegraphics[width=8cm]{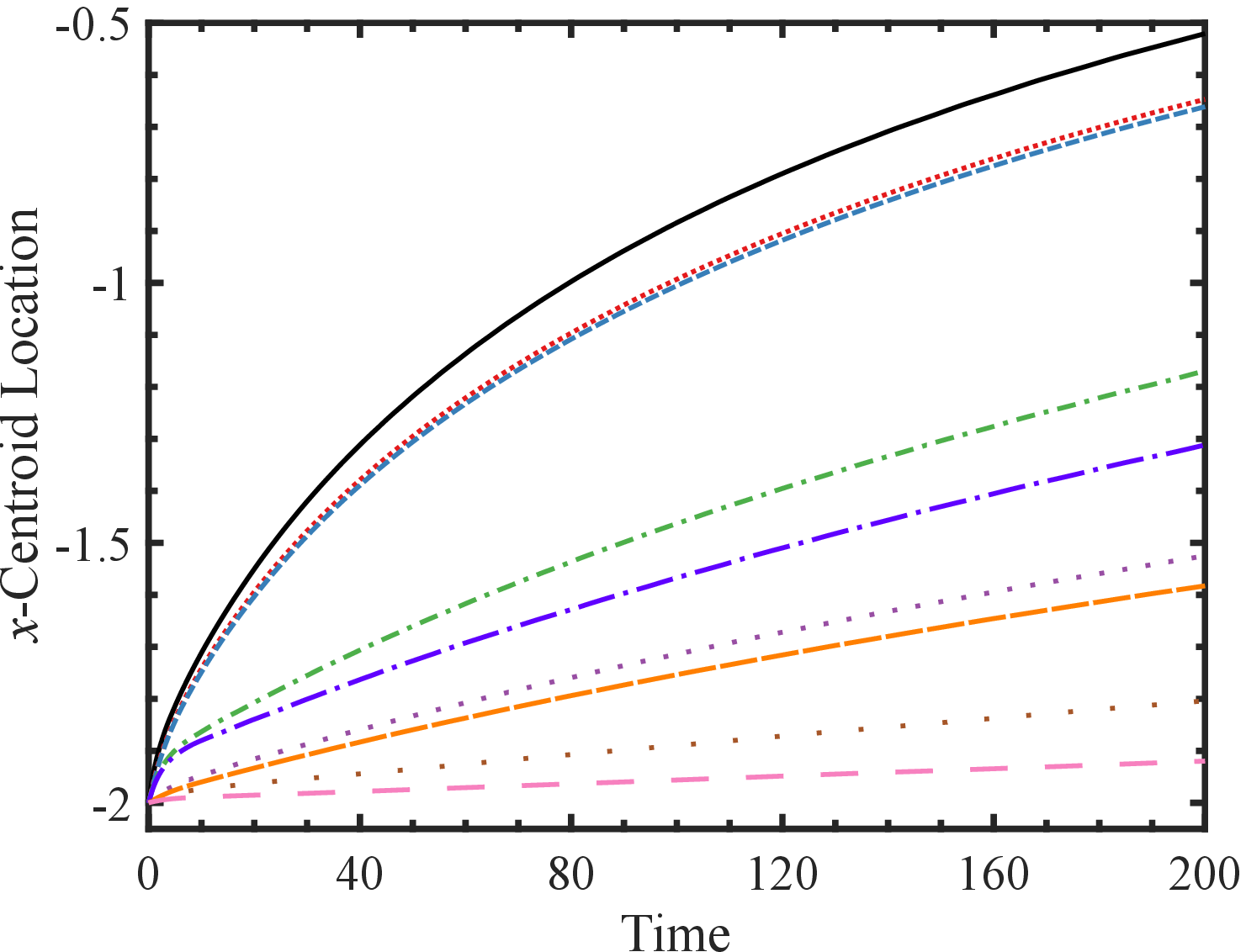}
		}		
		\subfigure[Deformation Parameter]{
			\includegraphics[width=8cm]{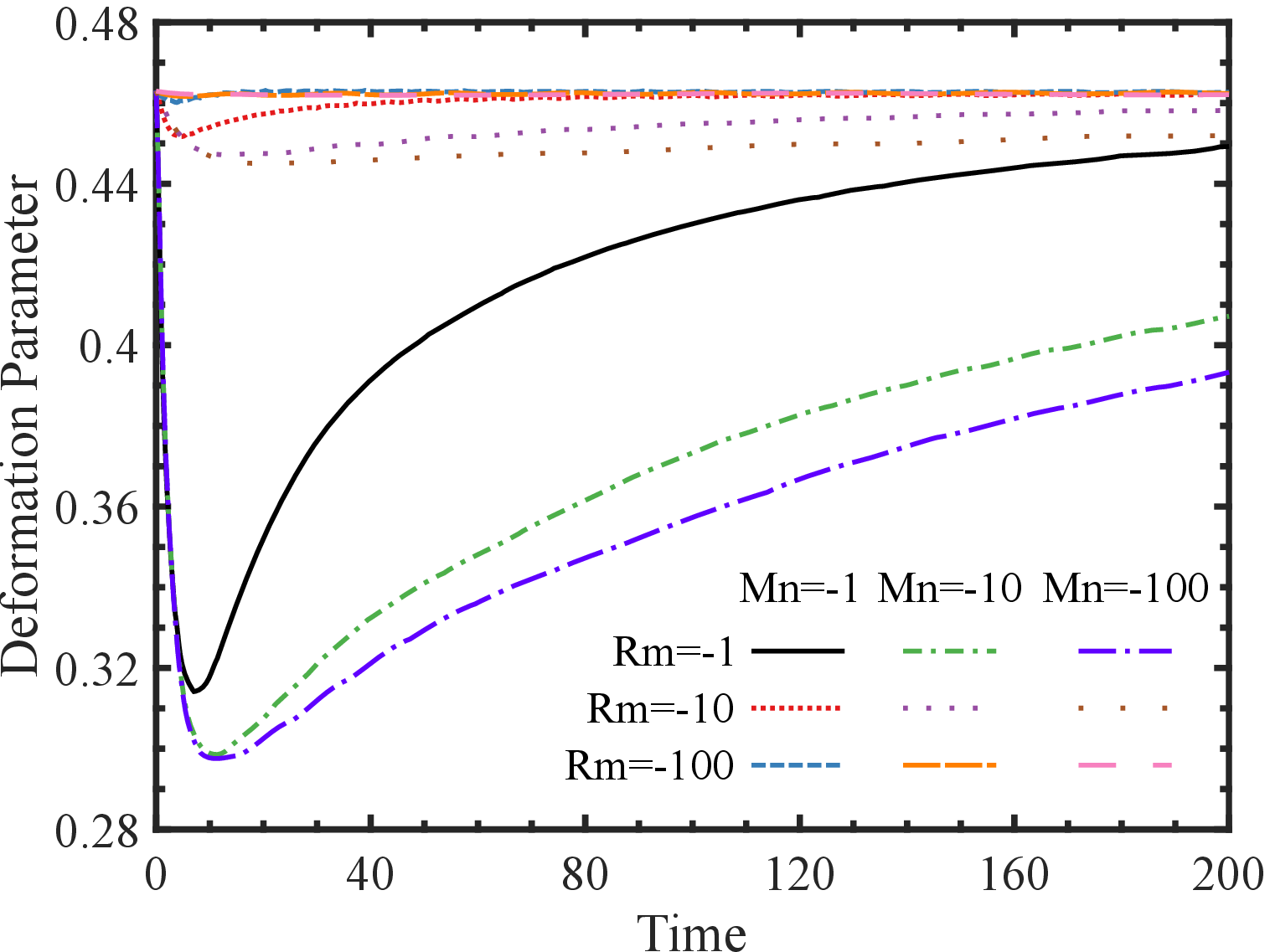}
		}		
		\caption{The evolution of the $x$-centroid and deformation parameter of the vesicle over time for various values of $\MN$ and $\RM$. 
			The combinations shown here correspond to those shown in Fig. \ref{fig:variableInterfaces}. The legend is common to both figures.}
		\label{fig:Variable_Results}
	\end{figure}

	The location of the $x$-centroid and the deformation parameter of the vesicle when exposed to this spatially varying magnetic field is shown in
	Fig. \ref{fig:Variable_Results}. It is clear that the fastest migration is achieved with small values of $\RM$ and $\MN$. Even in situations
	where the bulk-magnetic field effects are small, such as when $\MN=-100$, migration can occur due to the alignment energy. This is due to the fact
	that Eq. \eqref{eq:magneticRotEnergy} can be decreased by not only aligning the interface with the magnetic field, but also by 
	pushing the interface towards regions of lower magnetic field strength.
	
	The deformation parameter results mimic those seen in the spatially constant results. As the alignment strength increases, the vesicle becomes more deformed. 
	As the value of $\MN$ increases, this deformed state persists longer. This is due to the fact that it takes longer for the vesicle to migrate towards the 
	center of the domain when $\MN$ is large. This results in the vesicle remaining closer to the stronger and more compact magnetic field centered
	at $(-3.2,0)$.
	
	\begin{figure}
		\centering
		\subfigure[$\RM=-1$]{\includegraphics[width=8cm]{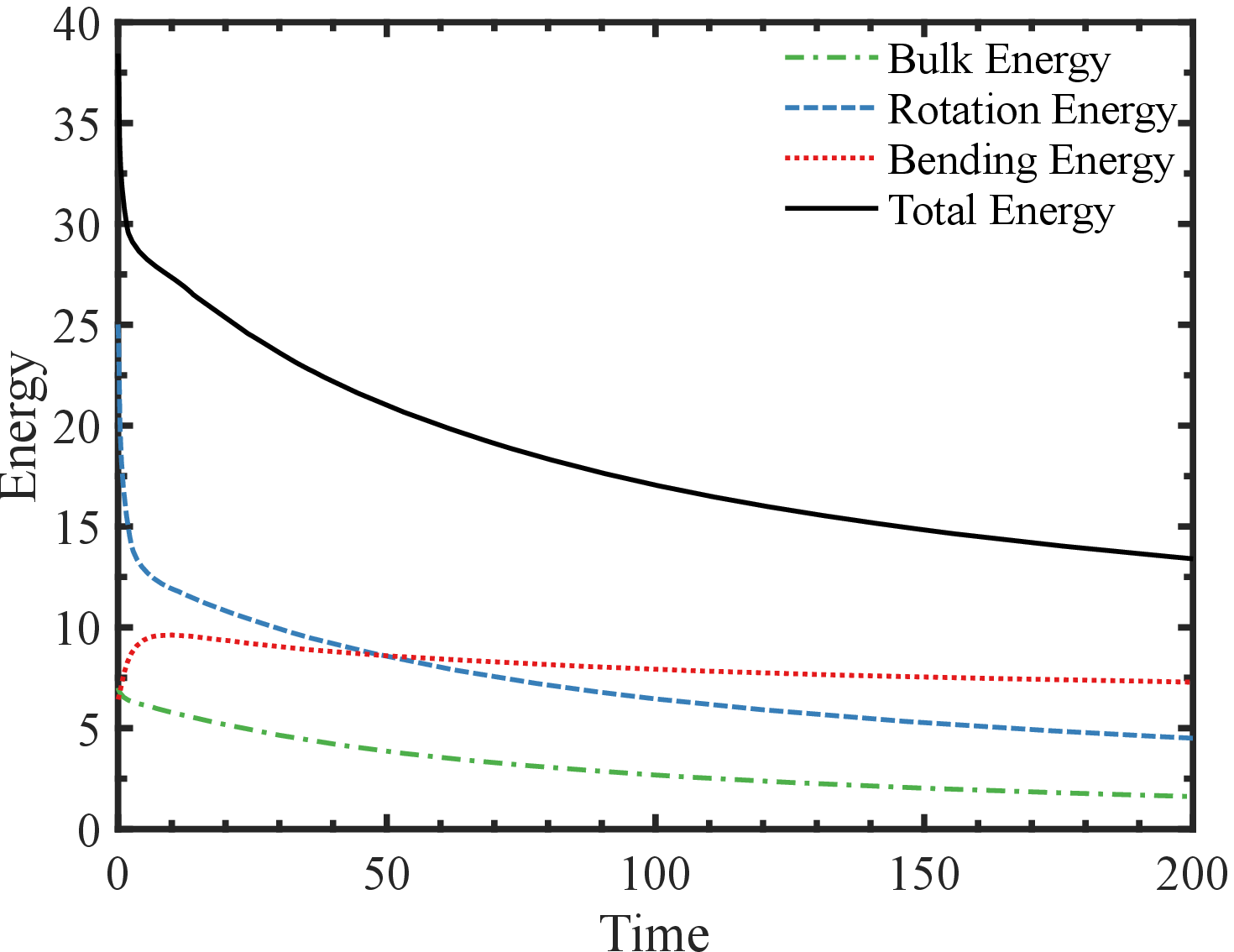}}
		\subfigure[$\RM=-10$]{\includegraphics[width=8cm]{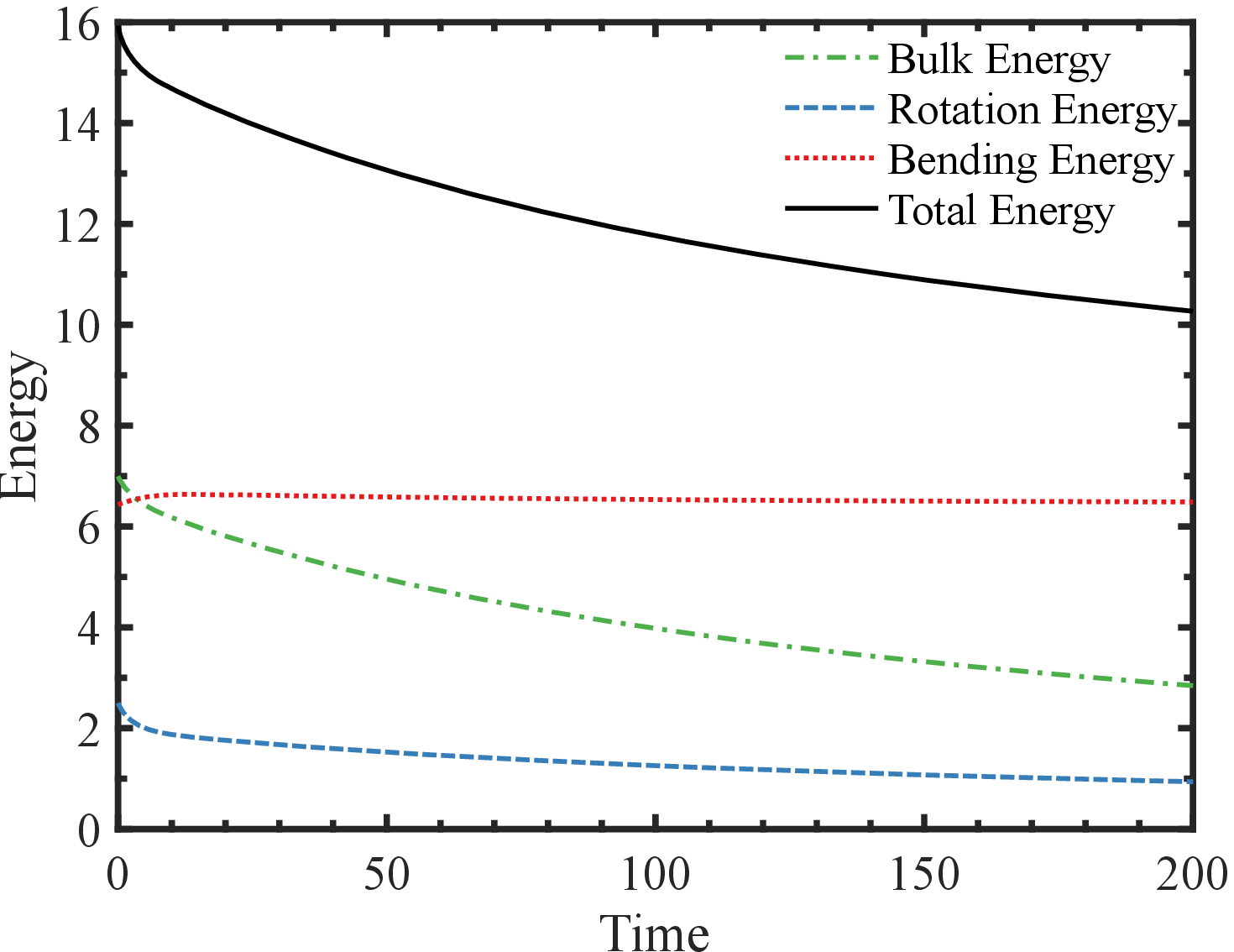}}
		\subfigure[$\RM=-100$]{\includegraphics[width=8cm]{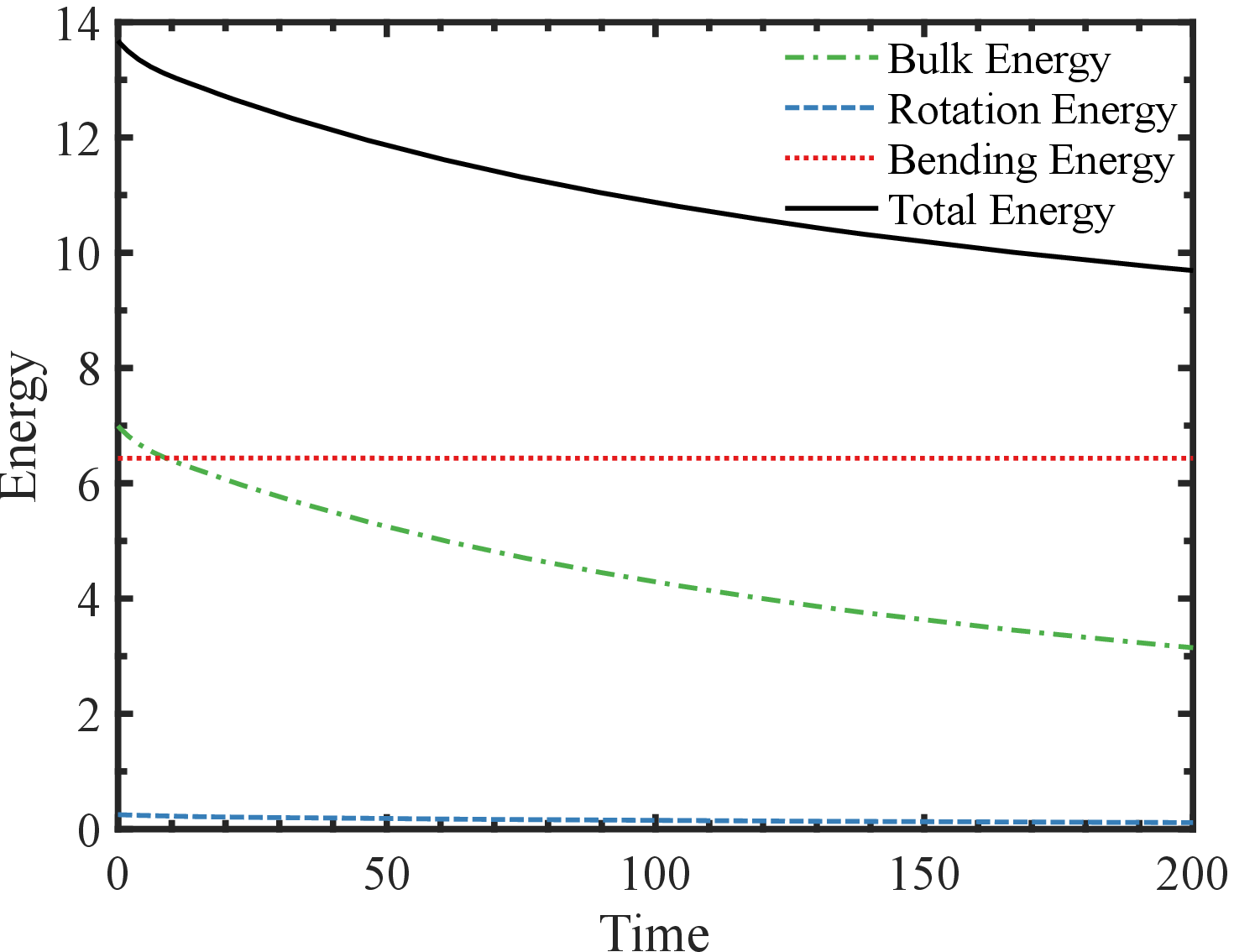}}
		\caption{The bending, bulk, rotation, and total energy for three different rotation strengths. The results assume $\MN=-10$.}
		\label{fig:Variable_energy}
	\end{figure}
	
	Finally, the energy of the system over time for a bulk constant of $\MN=-10$ and three alignment strengths, $\RM=-1$, $\RM=-10$, and $\RM=-100$ is shown
	in Fig. \ref{fig:Variable_energy}. As in Sec. \ref{sec:RM}, the magnetic energies are strictly decreasing over time. For the cases
	of strong magnetic field effects, particularly for $\MN=-10$ and $\RM=-1$, the bending energy increases above the initial value, and remains
	elevated throughout the simulation. It should be expected that as the vesicle moves towards the center of the domain, where the magnetic 
	field is weakest, the bending energy should have a larger influence.

\section{Conclusion}

	In this work a numerical model of vesicles in magnetic fields is presented. Based on the energy of the membrane, the interface forces
	due to magnetic rotation/alignment and the bulk magnetic energy are derived. These magnetic interface forces, in addition to the bending
	and tension forces of a vesicle, are used in conjunction with a level set description of the interface and a projection method for the fluid field
	to investigate the dynamics of a two-dimensional vesicle. The simulation is compared to the experimental results of Boroske and Helfrich,
	and good agreement is achieved. A systematic investigation of the influence of the rotation/alignment parameter, $\RM$,
	on the vesicle membrane is performed for spatially constant magnetic field. In general, there is a linear relationship between 
	$\RM$ and the amount of time it takes a vesicle to rotate through a particular angle. It was also demonstrated that if the angle
	between a vesicle and the magnetic field is $\pi/4$, then the alignment is not done through rotation, but by bulk deformation of the membrane.
	
	The movement of a vesicle in a spatially-varying magnetic field was also considered by placing a vesicle between two current-carrying wires.
	This magnetic field induced linear motion of the vesicle, with the rate of migration dependent on both the alignment parameter $\RM$ 
	and the bulk magnetic field parameter $\MN$. The particular nature of the underlying magnetic field induced deformations of the membrane,
	with the magnitude of these deformations depending on the particular parameter set.
	
	The use of magnetic fields opens up new possibilities for characterization and processing of not only liposome vesicle, but also other soft-matter multiphase systems such 
	as polymer vesicles or biological cells. For example, it is imagined that using the experimental equivalent to the simulations shown here 
	it could be possible to determine material properties
	such as the magnetic susceptibilities of the membrane molecules. This knowledge could then be used to design processing techniques, possibly in conjuncture with electric fields,
	to precisely control the dynamics of vesicles. Future work will explore these possibilities.

\section*{Acknowledgments}
	This work has been supported by the National Science Foundation through the Division of Chemical, Bioengineering, Environmental, and Transport Systems Grant \#1253739.

\begin{appendices}

\section{Calculus on Surfaces}
\label{App:Geometric}

	One issue with derivatives on surfaces is that operations require information of not only how 
	a function varies on the interface, but also how the interface itself varies. For this reason,
	some standard vector calculus identities may not hold. In this section the surface vector calculus 
	identities used to calculate the magnetic field force are derived.

%	Personal Note:
%Given a compact smooth surface $\Sigma\cup\mathbb{R}^3$, there exists a radius $r>0$ such that on
%	the set $S=\left\{\vec{x}\in \mathbb{R}^3:\textrm{dist}\left(x,\Sigma\right)<r\right\}$ we can solve the Eikonal 
%	equation $|\nabla\Psi|=1$ to get a function $\Psi:S\rightarrow \mathbb{R}$ such that $\Sigma=\Psi^{-1}(0)$
%	and $\nabla\Psi$ is the unit normal vector field for any level $\Psi^{-1}(c)$. Then in this formulation
%	we can see that the unit normal vector field is $\vec{n}=\nabla\Psi$ and thus $\nabla\times\vec{n}=\vec{0}$
%	on $\Sigma$. The number $r$, which is generally finite, is related to the radius of curvature of $\Sigma$.

	Let the interface be orientable with an outward unit normal $\vec{n}$. 
	Without loss of generality, it is assumed that the interface is described as the zero contour of 
	a function $\Psi$ such that $\Psi$ is the solution to the Eikonal equation, $|\nabla\Psi|=1$
	within a distance of $r$ to the interface, where $r$ depends on the curvature of the interface.
	With this assumption the normal is simply $\vec{n}=\nabla\Psi$. As the normal is now defined in a small region
	surrounding the interface, quantities such as the gradient of the unit normal, $\nabla\vec{n}$, are
	well-defined near the interface.
	
	The projection operator is 
	given by $\vec{P}=\vec{I}-\vec{n}\otimes\vec{n}$, or in component form $P_{ij}=\delta_{ij}-n_i n_j$, where $\delta_{ij}$ 
	is the Kronecker delta function. In this work, indices $i$ and $j$ are free indices and 
	while $p$, $q$, and $r$ are dummy indices. The projection operator is symmetric, $\vec{P}=\vec{P}^T$, and idempotent,
	\begin{align}
		\left[\vec{P}\vec{P}\right]_{ij}
			& =P_{ip}P_{pj}  \nonumber \\
			& =\left(\delta_{ip}-n_i n_p\right)\left(\delta_{pj}-n_p n_j\right) \nonumber \\
			& =\delta_{ip}\delta_{pj}-n_i n_p \delta_{pj} - n_p n_j \delta_{ip} + n_i n_p n_p n_j  \nonumber \\
			& = \delta_{ij} - n_i n_j - n_i n_j + n_i n_j \nonumber \\
			& = \delta_{ij} - n_i n_j = \left[\vec{P}\right]_{ij},
	\end{align}
	where $\left[\vec{v}\right]_i$ is the $i^{th}$ component of a vector $\vec{v}$, $\left[\vec{A}\right]_{ij}$ is the 
	$i,j$ compnent of a tensor $\vec{A}$, and 
	repeated indices indicate summation. 
	
	The generalized surface gradient function can be written as
	$\nabla_s\vec{A}=\left(\nabla\vec{A}\right)\vec{P}$, where $\vec{A}$ can be either a scalar, vector, or tensor field.~\cite{Fried2008,Gurtin1975,Napoli2012}
	For example, the surface gradient of a scalar field $a$ in component form would be written as 
	\begin{align}
		\left[\nabla_s a\right]_i
			=\left[\left(\nabla a\right)\vec{P}\right]_i
			=\dfrac{\partial a}{\partial x_p}P_{pi},
	\end{align}
	
	The surface gradient of a scalar field $a$ squared is
	\begin{align}
		\left[\nabla_s a^2\right]_i
			&=\left[\left(\nabla a^2\right)\vec{P}\right]_i=\dfrac{\partial a^2}{\partial x_p}P_{pi} \nonumber \\
			&=2a\dfrac{\partial a}{\partial x_p}P_{pi}=\left[2a \nabla_s a\right]_i.
	\end{align}
	For a vector field $\vec{v}$ the surface gradient would be 
	\begin{align}
		\left[\nabla_s \vec{v}\right]_{ij}
			=\left[\left(\nabla \vec{v}\right)\vec{P}\right]_{ij}
			=\dfrac{\partial v_i}{\partial x_p}P_{pj}.
	\end{align}
	The surface gradient of a vector dot product is
	\begin{align}
		\left[\nabla_s\left(\vec{v}\cdot\vec{w}\right)\right]_i&=\dfrac{\partial\left(v_p w_p\right)}{\partial x_q}P_{qi}\nonumber \\
			&= w_p \dfrac{\partial v_p}{\partial x_q}P_{qi} + v_p\dfrac{\partial w_p}{x_q}P_{qi} \nonumber \\
			&= \left[\vec{w}\cdot\nabla_s\vec{v}+\vec{v}\cdot\nabla_s\vec{w}\right]_i.
	\end{align}

	The surface divergence of any vector $\vec{v}$ can be written as 
	$\nabla_s\cdot\vec{v}=\textnormal{tr}\nabla_s\vec{v}=\vec{P}:\nabla\vec{v}$.~\cite{Fried2008}
	In component form this is written as
	\begin{align}
		\left[\nabla_s\cdot\vec{v}\right]
			=\left[\vec{P}:\nabla\vec{v}\right]
			=P_{pq}\dfrac{\partial v_p}{\partial x_q}.			
	\end{align}
	The surface divergence of a tensor field $\vec{A}$ is defined as\cite{Fried2008}
	\begin{align}
		\left[\nabla_s\cdot\vec{A}\right]_{i}=\left[\left(\nabla \vec{A}\right)\vec{P}\right]_i=\dfrac{\partial A_{ip}}{\partial x_q}P_{qp}.
	\end{align}
				
	The surface divergence of the projection operator is
	given by 
	\begin{align}
		\left[\nabla_s\cdot\vec{P}\right]_i
					& =\left[\left(\nabla \vec{P}\right)\vec{P}\right]_i
						=\dfrac{\partial P_{ip}}{\partial x_q}P_{qp} \nonumber \\
					&= \dfrac{\partial}{\partial x_q}\left(\delta_{ip}-n_i n_p\right)P_{qp}  \nonumber \\
					& =-\dfrac{\partial n_i}{\partial x_q}P_{qp}n_p -n_i\dfrac{\partial n_p}{\partial x_q}P_{qp} \nonumber \\
					& =\left[-\left(\nabla\vec{n}\right)\vec{P}\vec{n}-\vec{n}\nabla_s\cdot\vec{n}\right]_i
						=\left[-H \vec{n}\right]_i
	\end{align}
	due to the definition of total curvature, $H=\nabla_s\cdot\vec{n}$, and the fact that $\vec{P}\vec{n}=0$:
	\begin{align}
		\left[\vec{P}\vec{n}\right]_i
			& =P_{ip}n_p
				=\left(\delta_{ip}-n_i n_p\right)n_p \nonumber \\
			& =\delta_{ip}n_p - n_i n_p n_p=n_i-n_i
				=\left[0\right]_i.
	\end{align}	
	Let $a$ be a scalar field. The surface divergence of this scalar field times the projection operator is 
	\begin{align}
		\left[\nabla_s\cdot\left(a\vec{P}\right)\right]_i 
			&=\left[\left(\nabla\left(a\vec{P}\right)\right)\vec{P}\right]_i
				=\dfrac{\partial \left(a P_{ip}\right)}{\partial x_q}P_{qp} \nonumber \\
			& =\dfrac{\partial a}{\partial x_q}P_{ip}P_{qp}+a\dfrac{\partial{P_{ip}}}{\partial x_q}P_{qp} \nonumber \\
			&=\dfrac{\partial a}{\partial x_q}P_{qp}P_{pi}+a\dfrac{\partial{P_{ip}}}{\partial x_q}P_{qp} \nonumber \\
			&=\dfrac{\partial a}{\partial x_q}P_{qi}+a\dfrac{\partial{P_{ip}}}{\partial x_q}P_{qp}
				=\left[\nabla_s a-a H\vec{n}\right]_i.
	\end{align}	
	
	Next, consider the surface divergence of the tensor (outer) product of the unit normal $\vec{n}$ and any vector $\vec{v}$:
	\begin{align}
		\left[\nabla_s\cdot\left(\vec{n}\otimes\vec{v}\right)\right]_i
			& =\left[\left(\nabla\left(\vec{n}\otimes\vec{v}\right)\right)\vec{P}\right]_i 
				=\dfrac{\partial\left(n_i v_p\right)}{\partial x_q}P_{qp} \nonumber \\
			& = \dfrac{\partial n_i}{\partial x_q}P_{qp}v_p +n_i \dfrac{\partial v_p}{\partial x_q}P_{pq} 
				= \left[\left(\nabla_s\vec{n}\right)\vec{v} + \vec{n}\nabla_s\cdot\vec{v}\right]_i
	\end{align}
	
	Finally, consider the surface divergence of a scalar, the projection operator, and a vector,
	\begin{align}
		\left[\nabla_s\cdot\left(a\vec{P}\vec{v}\right)\right]
			&=\left[\vec{P}:\nabla\left(a\vec{P}\vec{v}\right)\right]
				=P_{pq}\dfrac{\partial\left(a P_{pr} v_r\right)}{\partial x_q} \nonumber \\
			& =P_{pq}\left(\dfrac{\partial a}{\partial x_q}P_{pr}v_r + a\dfrac{\partial P_{pr}}{\partial x_q}v_r +a P_{pr}\dfrac{\partial v_r}{\partial x_q}\right) \nonumber \\
			&=\dfrac{\partial a}{\partial x_q}P_{qr}v_r + a \dfrac{\partial P_{pr}}{\partial x_q} P_{pq} v_r + a \dfrac{\partial v_r}{\partial x_q} P_{qr} \nonumber \\
			& =\dfrac{\partial a}{\partial x_q}P_{qr}v_r + a \dfrac{\partial P_{rp}}{\partial x_q} P_{qp} v_r + a \dfrac{\partial v_r}{\partial x_q} P_{rq} \nonumber \\
			&=\left[\vec{v}\cdot\nabla_s a+a\left(\nabla_s\cdot\vec{P}\right)\cdot\vec{v}+a\nabla_s\cdot\vec{v}\right] \nonumber \\
			& =\left[\vec{v}\cdot\nabla_s a-aH \vec{n}\cdot\vec{v}+a\nabla_s\cdot\vec{v}\right]
	\end{align}

%&= \dfrac{\partial n_q}{\partial x_q}P_{pq} v_i +\dfrac{\partial v_i}{\partial x_q}P_{pq} n_q 
%				=\dfrac{\partial n_q}{\partial x_q}P_{pq} v_i 
%				=\left[\vec{v} \nabla_s\cdot\vec{n}\right]_i 
%				= \left[H \vec{v}\right]_i.			
	
\end{appendices}

%%%END OF MAIN TEXT%%%


%\balance


%%%REFERENCES%%%

\bibliographystyle{rsc} %the RSC's .bst file

\bibliography{magnetic}

\begin{thebibliography}{10}
\expandafter\ifx\csname url\endcsname\relax
  \def\url#1{\texttt{#1}}\fi
\expandafter\ifx\csname urlprefix\endcsname\relax\def\urlprefix{URL }\fi
\expandafter\ifx\csname href\endcsname\relax
  \def\href#1#2{#2} \def\path#1{#1}\fi

\bibitem{Kagan2010}
D.~Kagan, R.~Laocharoensuk, M.~Zimmerman, C.~Clawson, S.~Balasubramanian,
  D.~Kong, D.~Bishop, S.~Sattayasamitsathit, L.~Zhang, J.~Wang, {Rapid delivery
  of drug carriers propelled and navigated by catalytic nanoshuttles}, Small
  6~(23) (2010) 2741--2747.
\newblock \href {http://dx.doi.org/10.1002/smll.201001257}
  {\path{doi:10.1002/smll.201001257}}.

\bibitem{Toner2005}
M.~Toner, D.~Irimia, {Blood-on-a-chip}, Annual Review Of Biomedical Engineering
  7 (2005) 77--103.
\newblock \href {http://dx.doi.org/10.1146/annurev.bioeng.7.011205.135108}
  {\path{doi:10.1146/annurev.bioeng.7.011205.135108}}.

\bibitem{Xia2006}
N.~Xia, T.~P. Hunt, B.~T. Mayers, E.~Alsberg, G.~M. Whitesides, R.~M.
  Westervelt, D.~E. Ingber, {Combined microfluidic-micromagnetic separation of
  living cells in continuous flow}, Biomedical Microdevices 8~(4) (2006)
  299--308.
\newblock \href {http://dx.doi.org/10.1007/s10544-006-0033-0}
  {\path{doi:10.1007/s10544-006-0033-0}}.

\bibitem{Chen208}
P.~Chen, X.~Feng, W.~Du, B.-F. Liu, {Microfluidic chips for cell sorting},
  Frontiers In Bioscience-landmark 13 (2008) 2464--2483.
\newblock \href {http://dx.doi.org/10.2741/2859} {\path{doi:10.2741/2859}}.

\bibitem{Solovev2010}
A.~A. Solovev, S.~Sanchez, M.~Pumera, Y.~F. Mei, O.~G. Schmidt, {Magnetic
  control of tubular catalytic microbots for the transport, assembly, and
  delivery of micro-objects}, Advanced Functional Materials 20~(15) (2010)
  2430--2435.
\newblock \href {http://dx.doi.org/10.1002/adfm.200902376}
  {\path{doi:10.1002/adfm.200902376}}.

\bibitem{Tan2008}
Y.-C. Tan, Y.~Ho, A.~Lee,
  \href{http://dx.doi.org/10.1007/s10404-007-0184-1}{{Microfluidic sorting of
  droplets by size}}, Microfluidics and Nanofluidics 4~(4) (2008) 343--348.
\newblock \href {http://dx.doi.org/10.1007/s10404-007-0184-1}
  {\path{doi:10.1007/s10404-007-0184-1}}.
\newline\urlprefix\url{http://dx.doi.org/10.1007/s10404-007-0184-1}

\bibitem{DuBose2014}
J.~DuBose, X.~Lu, S.~Patel, S.~Qian, S.~{Woo Joo}, X.~Xuan, {Microfluidic
  electrical sorting of particles based on shape in a spiral microchannel},
  Biomicrofluidics 8~(1) (2014) 014101.

\bibitem{Wang2013}
G.~Wang, W.~Mao, R.~Byler, K.~Patel, C.~Henegar, A.~Alexeev, T.~Sulchek,
  {Stiffness Dependent Separation of Cells in a Microfluidic Device}, PLOS ONE
  8~(10) (2013) e75901.
\newblock \href {http://dx.doi.org/10.1371/journal.pone.0075901}
  {\path{doi:10.1371/journal.pone.0075901}}.

\bibitem{MacDonald2003}
M.~MacDonald, G.~Spalding, K.~Dholakia, {Microfluidic sorting in an optical
  lattice}, Nature 426~(6965) (2003) 421--424, 10.1038/nature02144.

\bibitem{Zhou2015}
R.~Zhou, C.~Wang, {Acoustic bubble enhanced pinched flow fractionation for
  microparticle separation}, Journal of Micromechanics and Microengineering
  25~(8) (2015) 084005.
\newblock \href {http://dx.doi.org/10.1088/0960-1317/25/8/084005}
  {\path{doi:10.1088/0960-1317/25/8/084005}}.

\bibitem{Dimova2009}
R.~Dimova, N.~Bezlyepkina, M.~D. Jordo, R.~L. Knorr, K.~A. Riske, M.~Staykova,
  P.~M. Vlahovska, T.~Yamamoto, P.~Yang, R.~Lipowsky, {Vesicles in electric
  fields: Some novel aspects of membrane behavior†}, Soft Matter 5 (2009)
  3201--3212.

\bibitem{dimova2007}
R.~Dimova, K.~A. Riske, S.~Aranda, N.~Bezlyepkina, R.~L. Knorr, R.~Lipowsky,
  {Giant vesicles in electric fields}, Soft matter 3~(7) (2007) 817--827.

\bibitem{Kolahdouz2015a}
E.~M. Kolahdouz, D.~Salac, {Dynamics of three-dimensional vesicles in dc
  electric fields}, Physical Review E 92~(1) (2015) 012302.
\newblock \href {http://dx.doi.org/10.1103/PhysRevE.92.012302}
  {\path{doi:10.1103/PhysRevE.92.012302}}.

\bibitem{salipante2014}
P.~F. Salipante, P.~M. Vlahovska, {Vesicle deformation in DC electric pulses},
  Soft Matter 10 (2014) 3386--3393.

\bibitem{schwalbe2011}
J.~T. Schwalbe, P.~M. Vlahovska, M.~J. Miksis, {Vesicle electrohydrodynamics},
  Physical Review E 83~(4) (2011) 046309.

\bibitem{mcconnell2013}
L.~C. McConnell, M.~J. Miksis, P.~M. Vlahovska, {Vesicle electrohydrodynamics
  in DC electric fields}, IMA Journal of Applied Mathematics 78~(4) (2013)
  797--817.

\bibitem{Barrett2005}
L.~M. Barrett, A.~J. Skulan, A.~K. Singh, E.~B. Cummings, G.~J. Fiechtner,
  {Dielectrophoretic manipulation of particles and cells using insulating
  ridges in faceted prism microchannels}, Analytical Chemistry 77~(21) (2005)
  6798--6804.
\newblock \href {http://dx.doi.org/10.1021/ac0507791}
  {\path{doi:10.1021/ac0507791}}.

\bibitem{Cummings2003}
E.~B. Cummings, A.~K. Singh, {Dielectrophoresis in microchips containing arrays
  of insulating posts: Theoretical and experimental results}, Analytical
  Chemistry 75~(18) (2003) 4724--4731.
\newblock \href {http://dx.doi.org/10.1021/ac0340612}
  {\path{doi:10.1021/ac0340612}}.

\bibitem{Vahey2008}
M.~D. Vahey, J.~Voldman, {An equilibrium method for continuous-flow cell
  sorting using dielectrophoresis}, Analytical Chemistry 80~(9) (2008)
  3135--3143.
\newblock \href {http://dx.doi.org/10.1021/ac7020568}
  {\path{doi:10.1021/ac7020568}}.

\bibitem{Riske2005}
K.~A. Riske, R.~Dimova, {Electro-deformation and poration of giant vesicles
  viewed with high temporal resolution}, Biophysical Journal 88~(2) (2005)
  1143--55.

\bibitem{BOROSKE1978}
E.~Boroske, H.~W., {Magnetic-anisotropy of egg lecithin membranes}, Biophysical
  Journal 24~(3) (1978) 863--868.

\bibitem{Rikken2014}
R.~S.~M. Rikken, R.~J.~M. Nolte, J.~C. Maan, J.~C.~M. van Hest, D.~A. Wilson,
  P.~C.~M. Christianen, {Manipulation of micro- and nanostructure motion with
  magnetic fields}, Soft Matter 10~(9) (2014) 1295--1308.
\newblock \href {http://dx.doi.org/10.1039/c3sm52294f}
  {\path{doi:10.1039/c3sm52294f}}.

\bibitem{Helfrich1973}
W.~Helfrich, {Lipid Bilayer Spheres - Deformation and Birefringence In
  Magnetic-fields}, Physics Letters A A 43~(5) (1973) 409--410.
\newblock \href {http://dx.doi.org/10.1016/0375-9601(73)90396-4}
  {\path{doi:10.1016/0375-9601(73)90396-4}}.

\bibitem{Helfrich1973a}
W.~Helfrich, {Elastic Properties of Lipid Bilayers - Theory and Possible
  Experiments}, Zeitschrift Fur Naturforschung C-a Journal of Biosciences C
  28~(11-1) (1973) 693--703.

\bibitem{Tenforde1988}
T.~S. Tenforde, L.~R. P., {Magnetic Deformation of Phospholipid-bilayers -
  Effects On Liposome Shape and Solute Permeability At Prephase
  Transition-temperatures}, Journal of Theoretical Biology 133~(3) (1988)
  385--396.
\newblock \href {http://dx.doi.org/10.1016/S0022-5193(88)80329-1}
  {\path{doi:10.1016/S0022-5193(88)80329-1}}.

\bibitem{Ozeki2000}
S.~Ozeki, H.~Kurashima, H.~Abe, {High-magnetic-field effects on liposomes and
  black membranes of dipalmitoylphosphatidylcholin: Magneotresponses in
  membrane potential and magnetofusion}, Journal of Physical Chemistry B
  104~(24) (2000) 5657--5660.
\newblock \href {http://dx.doi.org/10.1021/jp9934073}
  {\path{doi:10.1021/jp9934073}}.

\bibitem{Qiu1993}
X.~X. Qiu, P.~A. Mirau, C.~Pidgeon, {Magnetically Induced Orientation of
  Phosphatidylcholine Membranes}, Biochimica Et Biophysica Acta 1147~(1) (1993)
  59--72.
\newblock \href {http://dx.doi.org/10.1016/0005-2736(93)90316-R}
  {\path{doi:10.1016/0005-2736(93)90316-R}}.

\bibitem{Kiselev2008}
M.~A. Kiselev, T.~Gutberlet, A.~Hoell, V.~L. Aksenov, D.~Lombardo, {Orientation
  of the DMPC unilamellar vesicle system in the magnetic field: SANS study},
  Chemical Physics 345~(2-3) (2008) 181--184.
\newblock \href {http://dx.doi.org/10.1016/j.chemphys.2007.09.002}
  {\path{doi:10.1016/j.chemphys.2007.09.002}}.

\bibitem{Ye2015}
H.~Ye, A.~Curcuru, {Vesicle biomechanics in a time-varying magnetic field}, BMC
  Biophysics 8 (2015) 2.
\newblock \href {http://dx.doi.org/10.1186/s13628-014-0016-0}
  {\path{doi:10.1186/s13628-014-0016-0}}.

\bibitem{Seifert1997}
U.~Seifert, {Configurations of fluid membranes and vesicles}, Advances in
  Physics 46~(1) (1997) 13--137.

\bibitem{Saville1997}
D.~A. Saville, {Electrohydrodynamics: The Taylor-Melcher leaky dielectric
  model}, Annual Review of Fluid Mechanics 29~(1962) (1997) 27--64.

\bibitem{Melcher1969}
J.~Melcher, G.~Taylor, {Electrohydrodynamics: A review of the role of
  interfacial shear stresses}, Annual Review of Fluid Mechanics 1~(1) (1969)
  111--146.

\bibitem{Riske2006}
K.~A. Riske, R.~Dimova, {Electric pulses induce cylindrical deformations on
  giant vesicles in salt solutions}, Biophysical Journal 91~(5) (2006)
  1778--86.

\bibitem{needham1989electro}
D.~Needham, R.~Hochmuth, {Electro-mechanical permeabilization of lipid
  vesicles. Role of membrane tension and compressibility}, Biophysical Journal
  55~(5) (1989) 1001--1009.

\bibitem{Guo2013}
T.~Guo, S.~Wang, R.~Samulyak, {Sharp Interface Algorithm for Large Density
  Ratio Incompressible Multiphase Magnetohydrodynamic Flows}, 2013
  International Conference On Computational Science 18 (2013) 511--520.
\newblock \href {http://dx.doi.org/10.1016/j.procs.2013.05.215}
  {\path{doi:10.1016/j.procs.2013.05.215}}.

\bibitem{Ki2010}
H.~Ki, {Level set method for two-phase incompressible flows under magnetic
  fields}, Computer Physics Communications 181~(6) (2010) 999--1007.
\newblock \href {http://dx.doi.org/10.1016/j.cpc.2010.02.002}
  {\path{doi:10.1016/j.cpc.2010.02.002}}.

\bibitem{Tagawa2006}
T.~Tagawa, {Numerical simulation of two-phase flows in the presence of a
  magnetic field}, Mathematics and Computers In Simulation 72~(2-6) (2006)
  212--219.
\newblock \href {http://dx.doi.org/10.1016/j.matcom.2006.05.040}
  {\path{doi:10.1016/j.matcom.2006.05.040}}.

\bibitem{vlahovska2009electrohydrodynamic}
P.~M. Vlahovska, R.~S. Gracia, S.~Aranda-Espinoza, R.~Dimova,
  {Electrohydrodynamic model of vesicle deformation in alternating electric
  fields}, Biophysical Journal 96~(12) (2009) 4789--4803.

\bibitem{Carmo1976}
M.~Carmo, {Differential Geometry of Curves and Surfaces}, Prentice-Hall, 1976.

\bibitem{biben2003}
T.~Biben, C.~Misbah, {Tumbling of vesicles under shear flow within an
  advected-field approach}, Physical Review E 67~(3) (2003) 031908.

\bibitem{vlahovska2007}
P.~M. Vlahovska, R.~S. Gracia, {Dynamics of a viscous vesicle in linear flows},
  Physical Review E 75~(1) (2007) 016313.

\bibitem{Sohn2010}
J.~S. Sohn, Y.-H. Tseng, S.~Li, A.~Voigt, J.~S. Lowengrub, {Dynamics of
  multicomponent vesicles in a viscous fluid}, Journal of Computational Physics
  {229}~({1}) ({2010}) {119--144}.
\newblock \href {http://dx.doi.org/10.1016/j.jcp.2009.09.017}
  {\path{doi:10.1016/j.jcp.2009.09.017}}.

\bibitem{SCHOLZ1984}
F.~Scholz, B.~E., H.~W., {Magnetic-anisotropy of lecithin membranes - A new
  anisotropy susceptometer}, Biophysical Journal 45~(3) (1984) 589--592.

\bibitem{Tan2002}
C.~Tan, B.~Fung, G.~Cho, {Phospholipid bicelles that align with their normals
  parallel to the magnetic field}, {Journal of the American Chemical Society}
  {124}~({39}) ({2002}) {11827--11832}.
\newblock \href {http://dx.doi.org/10.1021/ja027079n}
  {\path{doi:10.1021/ja027079n}}.

\bibitem{seifert1999fluid}
U.~Seifert, {Fluid membranes in hydrodynamic flow fields: Formalism and an
  application to fluctuating quasispherical vesicles in shear flow}, The
  European Physical Journal B-Condensed Matter and Complex Systems 8~(3) (1999)
  405--415.

\bibitem{Napoli2010}
G.~Napoli, L.~Vergori,
  \href{http://stacks.iop.org/1751-8121/43/i=44/a=445207}{{Equilibrium of
  nematic vesicles}}, Journal of Physics A: Mathematical and Theoretical
  43~(44) (2010) 445207.
\newline\urlprefix\url{http://stacks.iop.org/1751-8121/43/i=44/a=445207}

\bibitem{Fried2008}
E.~Fried, M.~E. Gurtin, \href{http://dx.doi.org/10.1007/s00162-008-0083-4}{{A
  continuum mechanical theory for turbulence: a generalized
  Navier--Stokes-$\alpha$ equation with boundary conditions}}, Theoretical and
  Computational Fluid Dynamics 22~(6) (2008) 433--470.
\newblock \href {http://dx.doi.org/10.1007/s00162-008-0083-4}
  {\path{doi:10.1007/s00162-008-0083-4}}.
\newline\urlprefix\url{http://dx.doi.org/10.1007/s00162-008-0083-4}

\bibitem{Napoli2012}
G.~Napoli, L.~Vergori,
  \href{http://link.aps.org/doi/10.1103/PhysRevE.85.061701}{{Surface free
  energies for nematic shells}}, Physical Review E 85 (2012) 061701.
\newblock \href {http://dx.doi.org/10.1103/PhysRevE.85.061701}
  {\path{doi:10.1103/PhysRevE.85.061701}}.
\newline\urlprefix\url{http://link.aps.org/doi/10.1103/PhysRevE.85.061701}

\bibitem{Towers2009}
J.~D. Towers, {Finite difference methods for approximating Heaviside
  functions}, Journal of Computational Physics {228}~({9}) ({2009})
  {3478--3489}.
\newblock \href {http://dx.doi.org/10.1016/j.jcp.2009.01.026}
  {\path{doi:10.1016/j.jcp.2009.01.026}}.

\bibitem{Chang1996}
Y.~Chang, T.~Hou, B.~Merriman, S.~Osher, {A Level Set Formulation of Eulerian
  Interface Capturing Methods for Incompressible Fluid Flows}, Journal of
  Computational Physics 124~(2) (1996) 449--464.

\bibitem{Kolahdouz2015b}
E.~M. Kolahdouz, D.~Salac, {Electrohydrodynamics of Three-dimensional Vesicles:
  A Numerical Approach}, SIAM Journal on Scientific Computing 37~(3) (2015)
  B473--B494.
\newblock \href {http://dx.doi.org/10.1137/140988966}
  {\path{doi:10.1137/140988966}}.

\bibitem{Towers2008}
J.~D. Towers, {A convergence rate theorem for finite difference approximations
  to delta functions}, Journal of Computational Physics 227~(13) (2008)
  6591--6597.

\bibitem{Beblik1985}
G.~Beblik, S.~R. M., H.~W., {Bilayer Bending Rigidity of Some Synthetic
  Lecithins}, Journal De Physique 46~(10) (1985) 1773--1778.
\newblock \href {http://dx.doi.org/10.1051/jphys:0198500460100177300}
  {\path{doi:10.1051/jphys:0198500460100177300}}.

\bibitem{Velmurugan2016}
G.~Velmurugan, E.~M. Kolahdouz, D.~Salac,
  \href{http://www.sciencedirect.com/science/article/pii/S0045782516307423}{{Level
  Set Jet Schemes for Stiff Advection Equations: The SemiJet Method}}, Computer
  Methods in Applied Mechanics and Engineering 310 (2016) 233--251.
\newblock \href {http://dx.doi.org/10.1016/j.cma.2016.07.014}
  {\path{doi:10.1016/j.cma.2016.07.014}}.
\newline\urlprefix\url{http://www.sciencedirect.com/science/article/pii/S0045782516307423}

\bibitem{Seibold2012}
B.~Seibold, R.~R. Rosales, J.-C. Nave, {Jet schemes for advection problems},
  Discrete and Continuous Dynamical Systems - Series B 17~(4) (2012)
  1229--1259.

\bibitem{Vitkova2013}
V.~Vitkova, A.~G. Petrov,
  \href{http://www.sciencedirect.com/science/article/pii/B978012411516300005X}{{Chapter
  Five - Lipid Bilayers and Membranes: Material Properties}}, in: A.~Igli\v{c},
  J.~Genova (Eds.), {A Tribute to Marin D. Mitov}, Vol.~17 of {Advances in
  Planar Lipid Bilayers and Liposomes}, Academic Press, 2013, pp. 89--138.
\newblock \href {http://dx.doi.org/10.1016/B978-0-12-411516-3.00005-X}
  {\path{doi:10.1016/B978-0-12-411516-3.00005-X}}.
\newline\urlprefix\url{http://www.sciencedirect.com/science/article/pii/B978012411516300005X}

\bibitem{Salac2012}
D.~Salac, M.~J. Miksis,
  \href{http://journals.cambridge.org/article_S0022112012003801}{{Reynolds
  number effects on lipid vesicles}}, Journal of Fluid Mechanics 711 (2012)
  122--146.
\newblock \href {http://dx.doi.org/10.1017/jfm.2012.380}
  {\path{doi:10.1017/jfm.2012.380}}.
\newline\urlprefix\url{http://journals.cambridge.org/article_S0022112012003801}

\bibitem{RAMANUJAN1998}
S.~Ramanujan, P.~C.,
  \href{http://journals.cambridge.org/article_S0022112098008714}{{Deformation
  of liquid capsules enclosed by elastic membranes in simple shear flow: large
  deformations and the effect of fluid viscosities}}, Journal of Fluid
  Mechanics 361 (1998) 117--143.
\newblock \href {http://dx.doi.org/10.1017/S0022112098008714}
  {\path{doi:10.1017/S0022112098008714}}.
\newline\urlprefix\url{http://journals.cambridge.org/article_S0022112098008714}

\bibitem{Salac2011}
D.~Salac, M.~Miksis, {A level set projection model of lipid vesicles in general
  flows}, Journal of Computational Physics 230~(22) (2011) 8192--8215.

\bibitem{BRAGANZA1984}
L.~F. Braganza, B.~B. H., C.~T. J., M.~D., {The superdiamagnetic effect of
  magnetic-fields on one and 2 component multilamellar liposomes}, Biochimica
  Et Biophysica Acta 801~(1) (1984) 66--75.
\newblock \href {http://dx.doi.org/10.1016/0304-4165(84)90213-7}
  {\path{doi:10.1016/0304-4165(84)90213-7}}.

\bibitem{Tzirtzilakis2005}
E.~E. Tzirtzilakis,
  \href{http://scitation.aip.org/content/aip/journal/pof2/17/7/10.1063/1.1978807}{{A
  mathematical model for blood flow in magnetic field}}, Physics of Fluids
  17~(7).
\newblock \href {http://dx.doi.org/10.1063/1.1978807}
  {\path{doi:10.1063/1.1978807}}.
\newline\urlprefix\url{http://scitation.aip.org/content/aip/journal/pof2/17/7/10.1063/1.1978807}

\bibitem{Gurtin1975}
M.~E. Gurtin, A.~{Ian Murdoch}, \href{http://dx.doi.org/10.1007/BF00261375}{{A
  continuum theory of elastic material surfaces}}, Archive for Rational
  Mechanics and Analysis 57~(4) (1975) 291--323.
\newblock \href {http://dx.doi.org/10.1007/BF00261375}
  {\path{doi:10.1007/BF00261375}}.
\newline\urlprefix\url{http://dx.doi.org/10.1007/BF00261375}

\end{thebibliography}

\end{document}